\documentclass[preprint,superscriptaddress,nofootinbib]{revtex4-1}
\usepackage{graphicx}
\usepackage{dcolumn}
\usepackage{bm}
\usepackage{amsmath}
\usepackage{amsfonts} 
\usepackage{latexsym}
\usepackage{bbm}
\usepackage{color}
\usepackage{amssymb}
\usepackage{comment}
\usepackage{amsthm}
\usepackage{epsf}
\usepackage{epsfig}
\usepackage{caption3}
\usepackage{appendix}
\usepackage{cases}
\usepackage{subfigure}
\usepackage{multirow}
\usepackage{float}

\newcommand{\beq}{\begin {equation}}
\newcommand{\eeq}{\end   {equation}}
\newcommand{\bea}{\begin {eqnarray}}
\newcommand{\eea}{\end   {eqnarray}}
\newcommand{\baa}{\begin {array}   }
\newcommand{\eaa}{\end   {array}   }
\newcommand{\bit}{\begin {itemize} }
\newcommand{\eit}{\end   {itemize} }
\newcommand{\be }{\begin {equation}}
\newcommand{\ee }{\end   {equation}}

\newcommand{\wboson}{W^{\pm}}
\newcommand{\wbosonoff}{W^{\pm(*)}}
\newcommand{\zboson}{Z}

\newcommand{\fbinv}{\mathrm{fb}^{-1}}

\newcommand{\chboson}{H^{\pm}}

\begin{document}
\vspace*{-1.0truecm}
\title{Analysis of $W^\pm+4\gamma$  in the 2HDM Type-I at the LHC}
\author{Yan Wang}
\email{wangyan@imnu.edu.cn}
\affiliation{\small College of Physics and Electronic Information, Inner Mongolia Normal University, Hohhot 010022, PR China}
\author{A. Arhrib}
\email{aarhrib@gmail.com}
\affiliation{\small Abdelmalek Essaadi University, Faculty of Sciences and Techniques,  B.P. 2117 T{\'e}touan, Tanger, Morocco}
\author{R. Benbrik}
\email{r.benbrik@uca.ma}
\affiliation{\small Laboratoire de Physique Fondamentale et Appliqu{\'e}e de Safi, Facult{\'e} Polydisciplinaire de Safi, Sidi Bouzid, B.P. 4162, Safi, Morocco}
\author{M. Krab}
\email{mohamed.krab@usms.ac.ma}
\affiliation{\small Sultan Moulay Slimane University, Polydisciplinary Faculty, Research Team in Theoretical Physics and Materials (RTTPM), Beni Mellal, 23000, Morocco}
\author{B. Manaut}
\email{b.manaut@usms.ma}
\affiliation{\small Sultan Moulay Slimane University, Polydisciplinary Faculty, Research Team in Theoretical Physics and Materials (RTTPM), Beni Mellal, 23000, Morocco}
\author{S. Moretti}
\email{s.moretti@soton.ac.uk}
\affiliation{\small School of Physics and Astronomy, University of Southampton, Southampton SO17 1BJ,
United Kingdom}
\author{Qi-Shu Yan}
\email{yanqishu@ucas.ac.cn}
\affiliation{\small Center for Future High Energy Physics, Chinese Academy of Sciences, Beijing 100049, PR 
China}
\affiliation{\small School of Physics Sciences, University of Chinese Academy of Sciences, Beijing 100039,
PR China}
\begin{abstract}
{\footnotesize
   We analyse  a light charged Higgs boson in the 2-Higgs Doublet Model (2HDM) Type-I, when its mass satisfies the condition $M_{\chboson} < M_{t}+M_{b}$ and the  parameter space is consistent with theoretical requirements of
self-consistency as well as the latest experimental constraints from  Large Hadron Collider (LHC) and other data. Over such a  parameter space,
wherein the Standard Model (SM)-like state discovered at the LHC in 2012 is the heaviest CP-even state of the 2HDM, 
 it is found that the decay modes of the charged Higgs boson are dominated by $\chboson \rightarrow \wbosonoff h$. Furthermore, the light neutral Higgs boson $h$ dominantly decays into two photons. Under these conditions, we find that the production and decay process $ p p \to H^\pm h \to {W^\pm}^{(*)} h h \to \ell \nu_{\ell} + 4 \gamma$ ($\ell=e,\mu$) is essentially background free. However, since the $W^{\pm(*)}$ could be largely off-shell and the $h$ state is very light, so that both the lepton coming from the former and the photons coming from the latter could be rather soft, we  perform here a full Monte Carlo (MC) analysis at the detector level  demonstrating that such a $W^{\pm} + 4\gamma$  signal is very  promising, as it would be yielding significant excesses at the  LHC with an integrated luminosity of $L=$ 300 $\fbinv$ at both $\sqrt{s}= 13$ and  $14 ~\text{TeV}$. }
\end{abstract}

\maketitle

\section{Introduction}
The discovery of a 125 GeV scalar particle at the LHC represents  the last piece of the Standard Model (SM). Generally speaking, an agreement between the measured and predicted properties of this particle has been reached at a 2$\sigma$ level. But it is still interesting to  examine the possibility of whether the Higgs sector can include more (pseudo)scalar particles, which is quite natural in those new physics models with extra doublets or triplets. One of the typical feature of these new physics models  is that they predict one or more charged Higgs bosons. Thus, if a charged Higgs boson could be found at the LHC, it would be a clear evidence of  new physics Beyond the SM (BSM).

One of the simplest extensions of the SM is the 2HDM, which contains two complex Higgs  doublets. After  Electro-Weak Symmetry Breaking (EWSB), there are 3 Goldstone bosons, which are `eaten' by the $\wboson$ and  $\zboson$ bosons, and 5 degrees of freedom which incarnate 5 physical Higgs bosons. The latter are known as 2 neutral CP-even scalars ($h$ and $H$, with $M_h<M_H$), a CP-odd pseudoscalar ($A$) and two charged Higgs states  $\chboson$. 

In order to satisfy the stringent experimental constraints from  Flavour Changing Neutral Currents (FCNCs) at the tree level, typically, a $Z_2$ symmetry is introduced into the Yukawa sector such that each type of fermion only couples to one of the doublets of the 2HDM. Depending on the $Z_2$ charge assignment of the Higgs doublets, we can define four basic scenarios, known as (Yukawa) Types. In the Type-I scenario, all  fermions couple to the second Higgs doublet (labelled as $\Phi_2$) while the tree-level Yukawa couplings to the first Higgs doublet $\Phi_1$  are vanishing. In such a scenario, a light charged Higgs boson (say, lighter than $M_t-M_b$) is still allowed by even the most stringent bounds (from direct $H^\pm$ searches and $B$ physics 
measurements).

In a recent study \cite{Arhrib:2017wmo}, it was found that in the Type-I scenario of the 2HDM, the decay mode $\chboson \rightarrow \wbosonoff h$ could become the dominant one, in the so-called fermiophobic limit of the 2HDM Type-I. In these conditions, the main production and decay process of a $H^\pm$ state and, consequently, the emerging experimental signatures could be different from what the ATLAS and CMS Collaborations are looking for, which is the process $pp \to t \bar{t} \to W^\mp H^\pm b  \bar{b}$ with $H^\pm$ dominantly decaying into a pair of fermions ($\tau\nu$ and $cs$). At the same time, when also the $h$ state  is fermiophobic,  its decay modes into SM fermions could  be highly suppressed, which results into a large Branching Ratio (BR) for the mode $h\rightarrow \gamma\gamma$. therefore, in  \cite{Arhrib:2017wmo}, it was  found that the associated production process $pp \to H^\pm h$ could lead to a potentially detectable $\wboson+4\gamma$ final state. According to the parton level analysis in  \cite{Arhrib:2017wmo}, it was observed that this signature is almost background free and could have a large significance. Therefore, it is worth to examine whether this statement  is robust enough after taking into account parton shower, hadronisation, heavy flavour  decays and detector effects.

In this paper, like in  \cite{Arhrib:2017wmo}, we assume that the heaviest CP-even Higgs boson $H$ is the observed  SM-like Higgs boson, which properties are consistent with  measurements at the LHC. Furthermore, due to the constraints from EW precision tests, it is noted that the lighter Higgs boson $h$ can be lighter than 125 GeV. In such a parameter space, a light charged Higgs boson $H^\pm$  is thus  accompanied by a light Higgs boson $h$. We focus our collider phenomenology study on the signal process $p p \to H^\pm h \to {W^{\pm}}^{(*)} hh \to \ell\nu_\ell +  4 \gamma$ ($\ell=e,\mu$) and examine its feasibility at the LHC. It will eventually be found that, after taking into account theoretical and experimental constraints, there are points in the 2HDM Type-I parameter space which can be either discovered or ruled out already with the current integrated luminosity at the LHC and that, with the full Run 3 data set, or a tenth of the High-Luminosity LHC (HL-LHC) one \cite{Gianotti:2002xx,CidVidal:2018eel}, a definite statement on this BSM scenario can be made.

The paper is organised as follows. In Sect. \ref{sec_parameter}, we briefly describe the 2HDM and its Yukawa scenarios, then introduce a few  Benchmark Points (BPs) for our MC analysis which pass all present constraints, both theoretical and experimental. In Sect. \ref{sec_collider}, we perform a  detailed collider analysis of these BPs and examine the potential to discover the aforementioned signature of this 2HDM Type-I scenario. In Sect. \ref{sec_conclusion}, we present some conclusions.

\section{The 2HDM }
\label{sec_parameter}
The scalar sector of the 2HDM contains two complex $SU(2)$ doublets with hypercharge $Y = +1$,
\begin{eqnarray}
\Phi_a=\left(
\begin{array}{c}
\phi_a^+ \\
\phi_a^0
\end{array}
\right),\ \qquad  \left\langle 0|\Phi_a|0\right\rangle  = \left(
\begin{array}{c}
0\\v_a/\sqrt{2}
\end{array}
\right)\qquad ( a = 1,2),
\label{2hdm_doublets}
\end{eqnarray}
where 
$v_1$ and $v_2$ are the Vacuum Expectation Values (VEVs) of the neutral Higgs field components that break spontaneously the EW gauge symmetry to the Electro-Magnetic (EM) one, $SU(2)_L \otimes U(1)_Y \rightarrow U(1)_{\rm EM}$.
The most general $SU(2)_L\times U(1)_Y$ invariant scalar potential involving two Higgs doublets can be written as:
\begin{eqnarray}
V(\Phi_1,\Phi_2) &=& m_{11}^2 \Phi_1^\dagger\Phi_1+m_{22}^2\Phi_2^\dagger\Phi_2-[m_{12}^2\Phi_1^\dagger\Phi_2+{\rm h.c.}] \nonumber\\
&+& \frac{\lambda_1}{2}(\Phi_1^\dagger\Phi_1)^2
+\frac{\lambda_2}{2}(\Phi_2^\dagger\Phi_2)^2
+\lambda_3(\Phi_1^\dagger\Phi_1)(\Phi_2^\dagger\Phi_2)
+\lambda_4(\Phi_1^\dagger\Phi_2)(\Phi_2^\dagger\Phi_1) \nonumber\\
&+&\left\{\frac{\lambda_5}{2}(\Phi_1^\dagger\Phi_2)^2
+\big[\lambda_6(\Phi_1^\dagger\Phi_1)
+\lambda_7(\Phi_2^\dagger\Phi_2)\big]
\Phi_1^\dagger\Phi_2+{\rm h.c.}\right\}\,. \label{pot1}
\end{eqnarray}
The hermiticity of the potential requires all parameters to be real except $m_{12}^2$, $\lambda_5$, $\lambda_6$ and  $\lambda_7$. For simplicity, we will work with a CP-conserving scalar potential by choosing $m_{12}^2$ and $\lambda_{5, 6, 7}$ to be real.
Note that, in ${Z}_2$ symmetric models, terms that are proportional to $\lambda_6$ and $\lambda_7$ in the scalar potential are absent, to ensure the suppression of FCNCs at tree level (as already remarked upon).  

The Yukawa Lagrangian, which describes the interactions between the (pseudo)scalar fields and the fermion sector, is given as follows: 
\begin{eqnarray}
{\cal{L}}_Y &=& \bar{Q'}_L ( Y^{u}_1 {\tilde \Phi_{1}} + Y^{u}_2 {\tilde \Phi_{2}}) u'_{R}
+\bar{Q'}_L (Y^{d}_1 \Phi_{1} + Y^{d}_2 \Phi_{2}) d'_{R} 
+ \bar{L'_{L}} (Y^{l}_{1}\Phi_{1} + Y^{l}_{2}\Phi_{2}) l'_R + \rm{h.c.},
\label{yuk_III}
\end{eqnarray}
where $Q'_L$ and $L'_L$ are the weak isospin quark and lepton doublets,  $u'_R$ and $d'_R$ denote the right-handed quark singlets and  $Y_{1,2}^{u}$, $Y_{1,2}^{d}$ and $Y_{1,2}^{l}$ are coupling matrices in flavour space. 

The implementation of the aforementioned  discrete symmetry, depending on the ${Z}_2$ assignments, leads to  four Types of 2HDM: commonly denoted as Type-I, -II, -X and -Y. In the mass eigenstate basis, their treatment can be unified in the following form:
\begin{align}
-{\mathcal L}_Y^I=&+\sum_{f=u,d,\ell} \left[m_f \bar f f+\left(\frac{m_f}{v}\xi_h^f \bar f fh+\frac{m_f}{v}\xi_H^f \bar f fH-i\frac{m_f}{v}\xi_A^f \bar f \gamma_5fA \right) \right]\nonumber\\
&+\frac{\sqrt 2}{v}\bar u \left (m_u V \xi_A^u P_L+ V m_d\xi_A^d P_R \right )d H^+ +\frac{\sqrt2m_\ell\xi_A^\ell}{v}\bar\nu_L \ell_R h^+
+\text{h.c.},\label{Eq:Yukawa}
\end{align}
where $P_{L,R}=(1\pm \gamma_5)/2$ and $V$ denotes the Cabibbo-Kobayashi-Maskawa (CKM) matrix.

In our study,  like in \cite{Arhrib:2017wmo}, we choose to focus on Type-I, where only one doublet $\Phi_2$ couples to all fermions and thus the Higgs-fermion couplings are flavour diagonal in the fermion mass basis and depend only on two angles, $\alpha$ (parameterising the mixing between $h$ and $H$) and $\beta$ (which tangent is given by the ratio of the two VEVs), as shown in Tab~\ref{coup-typeI}.
\begin{table}[!h]
	\begin{center}
		\begin{tabular}{c|c|c|c} \hline\hline
			$\phi$  & $\xi^u_{\phi}$ &  $\xi^d_{\phi}$ &  $\xi^\ell_{\phi}$  \\   \hline
			$h$~ 
			& ~ $  \cos\alpha/ \sin\beta$~
			& ~ $  \cos\alpha/ \sin\beta$~
			& ~ $  \cos\alpha/ \sin\beta $~ \\
			$H$~
			& ~ $  \sin\alpha/ \sin\beta$~
            & ~ $  \sin\alpha/ \sin\beta$~
            & ~ $  \sin\alpha/ \sin\beta$~ \\
			$A$~  
			& ~ $  \cot \beta $~  
			& ~ $  -\cot \beta $~  
			& ~ $  -\cot \beta $~  \\ \hline \hline 
		\end{tabular}
	\end{center}
	\caption {Yukawa couplings of the $h$, $H$ and $A$ bosons to  quarks and leptons in the Type-I 2HDM.} 
\label{coup-typeI}
\end{table}

\subsection{Constraints on the 2HDM}
\label{sec3}
There are certain theoretical restrictions and experimental constraints on the scalar potential that have to be imposed in order to obtain a viable realisation of the 2HDM. We note that both theoretical consistency and experimental data have already limited the parameter space of the 2HDM. 

In our study, we consider the following theoretical constraints.
\begin{itemize}
	\item[(1)]  Perturbativity is not invalidated in the Higgs sector, so long that 
	$\lambda_i < 4\pi$ ($i=1,\ldots,5$). 
	
	\item[(2)] The $S$-matrix satisfies all relevant tree-level unitarity constraints,
	which implies that the quartic couplings $\lambda_i$ satisfy the following relations:
	\cite{Akeroyd:2000wc}
	\begin{eqnarray}
	&&
	3(\lambda_1+\lambda_2)\pm\sqrt{9(\lambda_1-\lambda_2)^2+4(2\lambda_3+\lambda_4|)^2}
	<16 \pi,  \nonumber \\
	&&
	\lambda_1+\lambda_2\pm\sqrt{(\lambda_1-\lambda_2)^2+4|\lambda_5|^2}
	<16\pi,  \nonumber \\
	&&
	\lambda_1+\lambda_2\pm\sqrt{(\lambda_1-\lambda_2)^2+4|\lambda_5|^2}
	<16 \pi,  \nonumber \\
	&&\lambda_3+2\lambda_4\pm 3 | \lambda_5|<8\pi, \nonumber\\
	&& \lambda_3\pm\lambda_4 < 8\pi,\nonumber\\
	&& \lambda_3\pm|\lambda_5|<8\pi.
	\label{eq:PertBounds}
	\end{eqnarray}
	\item[(3)] The scalar potential is finite at large
	field values and contains no flat directions, which translate
	into the bounds \cite{Gunion:2002zf}:
	\begin{eqnarray}
	\lambda_{1,2}>0,  \,\,
	\lambda_3>- \sqrt{\lambda_1\lambda_2}, \,\,
	\lambda_3+\lambda_4-|\lambda_5|> - \sqrt{\lambda_1\lambda_2}.
	\label{eq:VacStabBounds}
	\end{eqnarray}
\end{itemize}

On the experimental side, we consider the following constraints.
\begin{itemize}
	
	\item[(4)] Limits from the EW oblique parameters $S$, $T$ and $U$ \cite{Peskin92}, for which we check their consistency at 95\% Confidence Level (CL) with the following measurements \cite{ParticleDataGroup:2014cgo}:
	\begin{equation}
	S = 0.05 \pm 0.11,\quad T = 0.09 \pm 0.13,\quad U = 0.01 \pm 0.11.
	\end{equation}
	
	\item[(5)] To study the effects of LEP, TeVatron and  LHC data affecting the Higgs sector, we have considered both exclusions from nil searches for Higgs boson companions, via HiggsBounds-5.9.0 \cite{Bechtle:2020pkv}, and measurements of the SM-like Higgs boson properties, via  HiggsSignals-2.6.0 \cite{Bechtle:2020uwn} (for which we have enforced  a  best fit at $95.5\%$ CL, which corresponds to $\Delta \chi^2 (\chi^2 - \chi^2_{\rm min}) \leq 5.99$).
	
	\item[(6)]  Constraints from  $B$ physics observables, which 
	give rise to bounds on the parameter space of the 2HDM,  as per the measured values in Tab.~\ref{Tab:ExpResult} (where we also give the corresponding SM predictions).\\
    	\begin{table}[h!]
    	\begin{center}
    		\begin{tabular}{|c|c|c|}
    			\hline 
    			\hline Observable 		&  Experimental result & SM prediction \\ \hline 
    			\,\,BR$(B\to X_s \gamma)$ & $(3.32 \pm 0.15) \times 10^{-4}$~\cite{HFLAV:2016hnz}& $(3.34 \pm 0.22) \times 10^{-4}$ \\ \hline
    			~~~~BR$(B_s \to \mu^+ \mu^-)$ & $(3.0 \pm 0.6 \pm 0.25) \times 10^{-9}$~\cite{LHCb:2017rmj}& $(3.54 \pm 0.27) \times 10^{-9}$ \\\hline
    			BR$(B_d \to \tau \nu)$ & $(1.06 \pm 0.19) \times 10^{-4}$~\cite{HFLAV:2016hnz} &$(0.82 \pm 0.29) \times 10^{-4}$ \\
    			\hline 
    			\hline
    	\end{tabular}\end{center}
    	\caption{Experimental results and  SM predictions for selected flavour observables.}
    	\label{Tab:ExpResult}
    \end{table}
\end{itemize}
(Note that we have used SuperIso v4.1 \cite{Mahmoudi:2008tp} to compute the exclusions from  flavour physics observables and 2HDMC \cite{Eriksson:2009ws} to check the theoretical constraints as well as the parameters $S$, $T$ and $U$.)


\subsection{Parameter space scans}

In this work, we concentrate on the scenario in which the $h$ state is  fermiophobic, which occurs near the alignment limit $\sin(\beta-\alpha)\sim 0$. In such a limit, all fermionic decays of the lightest CP-even Higgs boson are suppressed, so that $h\to \gamma\gamma$ can become significant. In the SM, the $\gamma\gamma$ decay of the Higgs boson is generated by the dominant $\wboson$ loop  and the subdominant top quark one, which have opposite signs and thus cancel one another somewhat. In the 2HDM, the additional
$\chboson$ loop also contributes. In the Type-I case, one has the following coupling dependencies for the lightest CP-even state: $hW^{+}W^{-}\sim \sin(\beta-\alpha)$,  $hq\bar{q}\sim \cos\alpha/\sin\beta$ while the $hH^{+}H^{-}$ vertex  is given by the  parameters of the scalar potential.   Since, for  fermionic loops, the coupling is proportional to $\cos\alpha$ and 
\begin{equation}
    \cos\alpha = \sin\beta\sin(\beta-\alpha) + \cos\beta\cos(\beta-\alpha),
\end{equation}
when $\sin(\beta-\alpha)$ is negative and $\cos(\beta-\alpha)$ is positive, $\cos\alpha$ will be cancelled for a particular $\tan\beta$, which is when $h$ becomes fermiophobic and $h\to \gamma\gamma$ is enhanced because the aforementioned cancellation no longer occurs. 

A numerical scan of the 2HDM Type-I parameters was performed in Ref.~\cite{Arhrib:2017wmo}, which satisfied all theoretical and experimental constraints mentioned above.  Based on the same scan, in this work, we propose the following 14 BPs given in  Tab.~\ref{t_bp_para}. There are three comments to make on these 14 BPs.
\begin{itemize}
\item The mass of the charged Higgs boson can vary from 91.49 GeV to 168.69 GeV. The CP-even Higgs boson $h$ is always lighter than 125 GeV and lighter than the ${H^\pm}$ state. The $W^\pm$ boson from the charged Higgs boson decaying via $H^\pm \to {W^\pm}^{(*)} h$ could be either on-shell or off-shell. If it is off-shell, the charged lepton emerging from it  might be soft (as already remarked upon), like in BP5-BP10. 
\item The parameter $\sin(\beta-\alpha)$ is constrained by the SM-like Higgs boson data from the LHC via the measurements of  $H \to W^\pm W^{\mp(*)}$ and $Z Z^*$, since the BRs of these modes are proportional to $\cos^2(\beta-\alpha)$. Current LHC Higgs data essentially demand that $ |\sin^2(\beta - \alpha) = 1 - \cos^2(\beta-\alpha) | < 0.10$, which leads to a range of $|\sin(\beta-\alpha)| < 0.3$ or so. 
\item For these BPs, the main production process of a charged Higgs boson is $pp \to H^\pm h$, which cross section can be up to one order of  magnitude larger than those of $pp \to H^\pm A$ and $pp \to H^\pm H^\mp$, which are alternative discovery modes in this region of 2HDM Type-I parameter space \cite{Arhrib:2021xmc, Arhrib:2016wpw, Bahl:2021str}. Therefore, in the present analysis, we will focus on the $W^\pm + 4\gamma$ signature stemming from the $pp \to H^\pm h$  production process only. 
\end{itemize}
Thus, since our signal is given by 
 $p p \to H^\pm h \to {W^\pm}^{(*)}hh \to \ell \nu_\ell + 4 \gamma$,  
the dominant background processes are $W^\pm +4j0\gamma$, $W^\pm+3j1\gamma$, $W^\pm+2j2\gamma$, $W^\pm+1j3\gamma$ and $W^\pm+0j4\gamma$, where a jet has a certain probability to fake a photon.

\begin{table}
\begin{center}
\begin{tabular}{| c| c| c| c| c| c| c| c| c|}
\hline
&$M_h$&$M_A$&$M_{H^{\pm}}$&$\sin(\beta-\alpha)$&$\tan\beta$&$m_{12}^{2}$&$\sigma_{13}(W+4\gamma)$ [fb]&$\sigma_{14}(W+4\gamma)$ [fb]\\
\hline
BP1&25.57&72.39&111.08&$-0.074$&13.58&11.97&101.40&112.55\\
\hline
BP2&35.12&111.24&151.44&$-0.075$&13.32&16.66&167.75&186.20\\
\hline
BP3&45.34&162.07&128.00&$-0.136$&7.57&80.96&10.76&11.93\\
\hline
BP4&53.59&126.09&91.49&$-0.127$&8.00&51.16&27.05&29.88\\
\hline
BP5&63.13&85.59&104.99&$-0.056$&18.09&190.24&179.31&198.61\\
\hline
BP6&65.43&111.43&142.15&$-0.087$&11.52&325.36&174.49&194.30\\
\hline
BP7&67.82&79.83&114.09&$-0.111$&8.94&326.32&177.72&197.23\\
\hline
BP8&69.64&195.73&97.43&$-0.111$&8.86&357.10&196.04&217.18\\
\hline
BP9&73.18&108.69&97.34&$-0.122$&8.06&594.64&193.56&214.57\\
\hline
BP10&84.18&115.26&148.09&$-0.067$&14.82&473.88&61.92&68.98\\
\hline
BP11&68.96&200.84&155.40&$-0.112$&8.64&531.46&62.02&69.14\\
\hline
BP12&71.99&91.30&160.10&$-0.104$&9.74&472.22&58.99&65.80\\
\hline
BP13&74.09&102.49&163.95&$-0.092$&10.56&503.74&55.58&62.04\\
\hline
BP14&81.53&225.76&168.69&$-0.101$&9.75&501.29&51.85&57.91\\
\hline
\end{tabular}
\end{center}
\caption{Input  parameters and the Leading Order (LO) cross sections at the parton-level with $\sqrt{s}=13(14)$ TeV for each BP are presented. All masses are in GeV and recall that $M_{H}$ = 125 GeV.}\label{t_bp_para} 
\end{table}

\section{Collider Phenomenology}\label{sec_collider}
In this section, we present a detailed MC analysis at a detector level, including both signal and background events. 
\subsection{Event generation}
Here we briefly describe MC event generation.
\begin{itemize}
\item  We use MadGraph5\_aMC@NLO  v2.8.2 \cite{Alwall:2014hca} (MG) to compute the cross sections and generate both signal and background events at parton level. We have adopted the following kinematic cuts  
(in pseudorapidity, transverse momentum and Missing $E_T$ (MET), where $E_T$ is the transverse energy (or momentum)) in order to improve the efficiency of the MC event generation
\begin{eqnarray}
|\eta(l,j,\gamma)|<2.5, \quad p_T(j,\gamma,l)>10 ~\text{GeV}, \quad \Delta R(l,j,\gamma)>0.5,  \quad \textrm{MET} > 5 ~\text{GeV},
\end{eqnarray}
where $j$ refers here to parton. The signal events are generated at LO, the cross sections for each BPs at the LHC with $\sqrt{s}=13 (14)$ TeV are listed in the last two column of Tab. \ref{t_bp_para}. The backgrounds are treated at LO, but this apparent inconsistency will become irrelevant once selection cuts are implemented, as the signal will be proven to be essentially backgound free for all  BPs.

\item After generating both signal and background events at the parton level, we pass them to Pythia v8  \cite{Sjostrand:2006za} to simulate initial and final state radiation (i.e., the QED and QCD emission), parton shower,  hadronisation and heavy flavour decays. 

\item We use Delphes v3.4.2  \cite{deFavereau:2013fsa} to simulate the detector effects. 
For each event, we cluster final particles into jets and we adopt the anti-$k_t$ jet algorithm \cite{Cacciari:2008gp} with jet parameter $\Delta R=0.5$ in the FastJet package \cite{Cacciari:2011ma}\footnote{Results obtained from the $k_t$ \cite{Catani:1993hr} or Cambridge-Aachen \cite{Dokshitzer:1997in,Wobisch:1998wt} algorithms are very similar.}. Following the  ATLAS analysis of \cite{TheATLAScollaboration:2015lks}, we will take the  fake rate as  0.001, which describes  the probability to mistag a jet as a photon. 
\end{itemize}
Notice that, in the following, we will present event rates corresponding to an LHC energy of $\sqrt s=13$ TeV and 14 TeV and integrated luminosity of $L= 300$ fb$^{-1}$. 

\subsection{Event reconstruction}
\begin{figure}[ht]
     \begin{center}    
 \includegraphics[height=7cm]{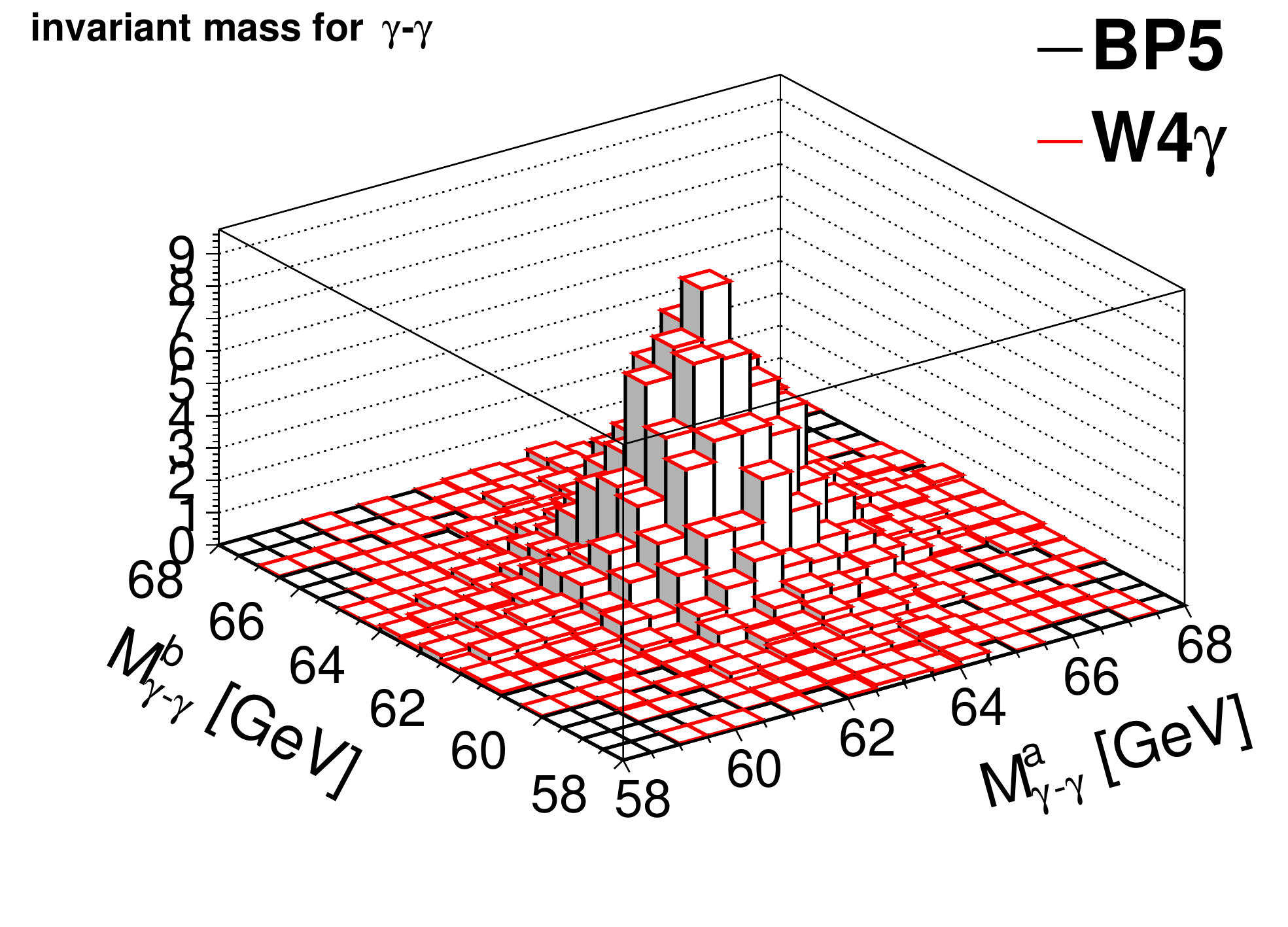} 
  \end{center}  
  \caption{The 3D $M_{\gamma\gamma}$ distribution for the two$\gamma\gamma$  pairings of the BP5 signal. }\label{f_mh1}
\end{figure}

The  mass $M_{h}$ can be reconstructed on an event-by event basis by pairing the four photons into two pairs by minimising the following $\chi^2$:
\begin{equation}
    \chi^2=(M_{\gamma\gamma}^{1}-M_{h})^2 +(M_{\gamma\gamma}^{2}-M_{h})^2\,.
\end{equation}
Obviously, there are 3 combinatorics for each event. When the combination which minimises the $\chi^2$ is found, we label the larger invariant mass of the pair of two photons as $M_{\gamma\gamma}^{1}$ and the other one is then labelled as $M_{\gamma\gamma}^{2}$.  The distributions of these two reconstructed masses of $h$ for, e.g., BP5 are displayed in Fig.~\ref{f_mh1}, where $M_{\gamma\gamma}^{1}$ and $M_{\gamma\gamma}^{2}$ are close to $M_{h}$ at the same time. 

\begin{figure}[ht]
     \begin{center}    
 \includegraphics[height=7cm]{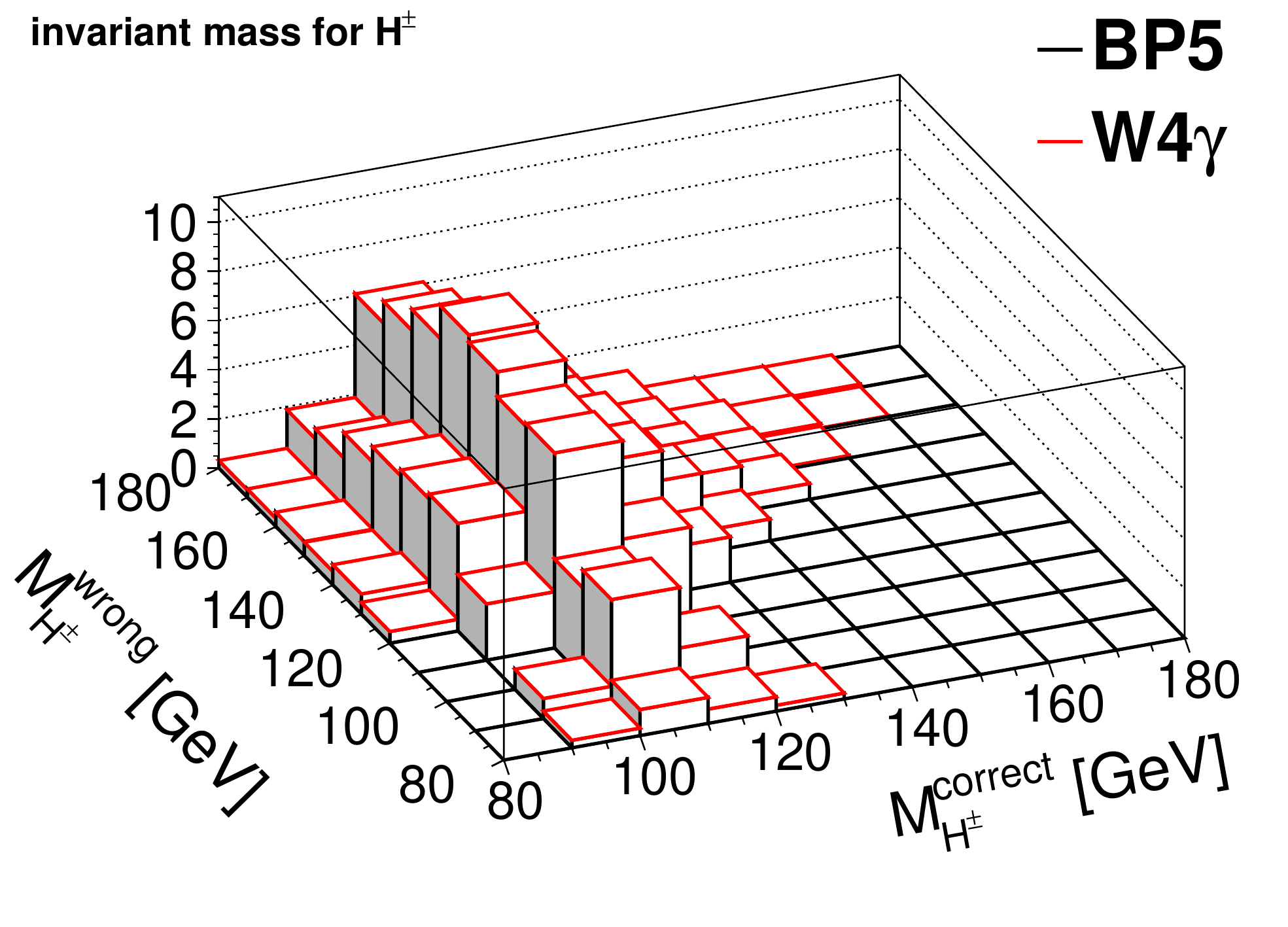} 
  \end{center}  
  \caption{The 3D $M_{H^{\pm}}$ distribution for the two  different combinations of the BP5 signal. }\label{f_mch}
\end{figure} 

The distribution of the reconstructed mass of the charged Higgs boson is shown in Fig.~\ref{f_mch}.  Since there is  missing energy, we use the standard method as the $W^\pm$ boson reconstruction. There are then two possible candidates for the light Higgs state, one is produced in the charged Higgs boson decay and the other is produced in association with it. Thus, we obtained two possible $H^\pm$ masses. As shown in Fig.~\ref{f_mch}, the correct one is rather sharp, while the wrong one is more dispersed. 

\begin{figure}[htp]
    \begin{center}  
\includegraphics[height=5.5cm]{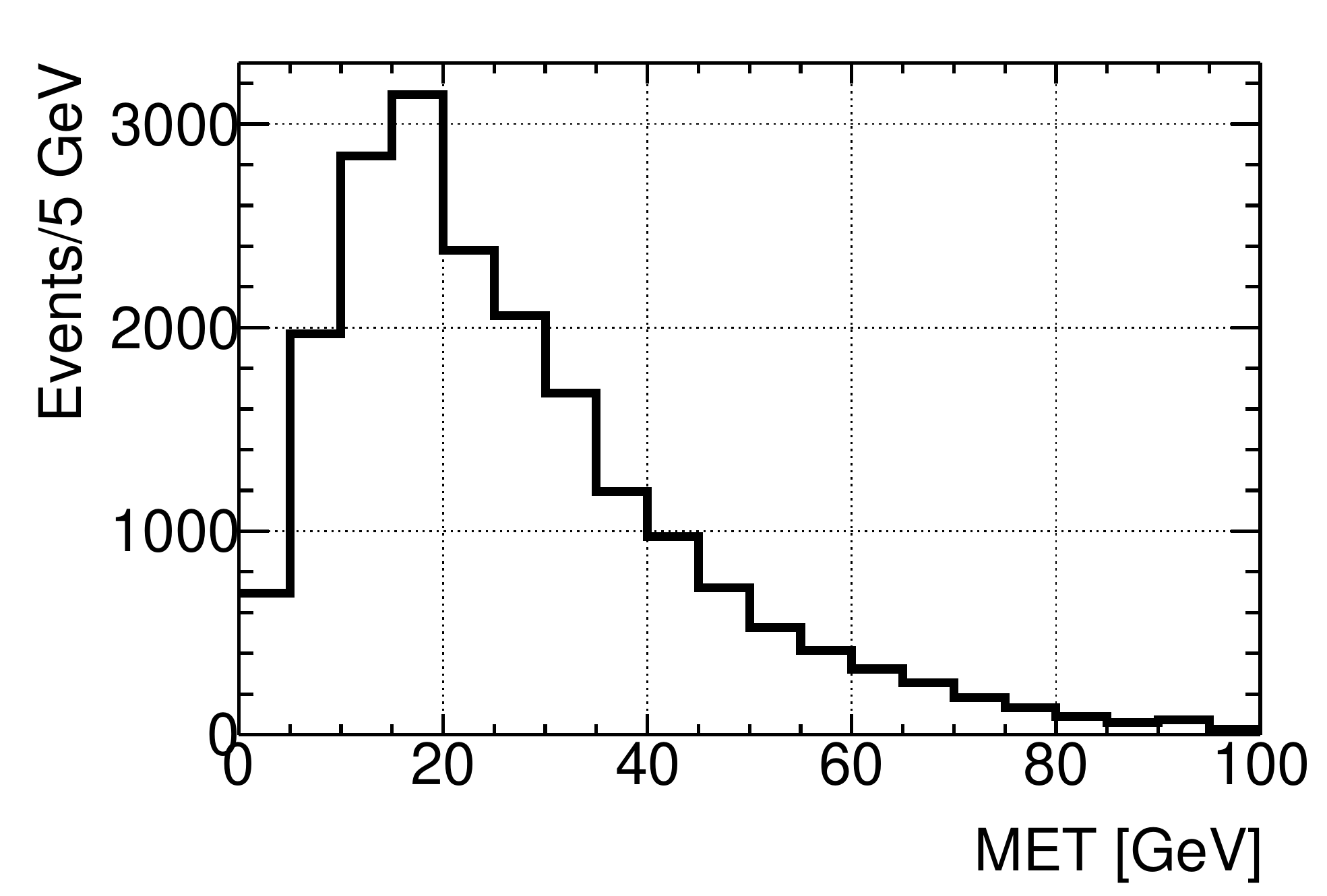} 
 \includegraphics[height=5.5cm]{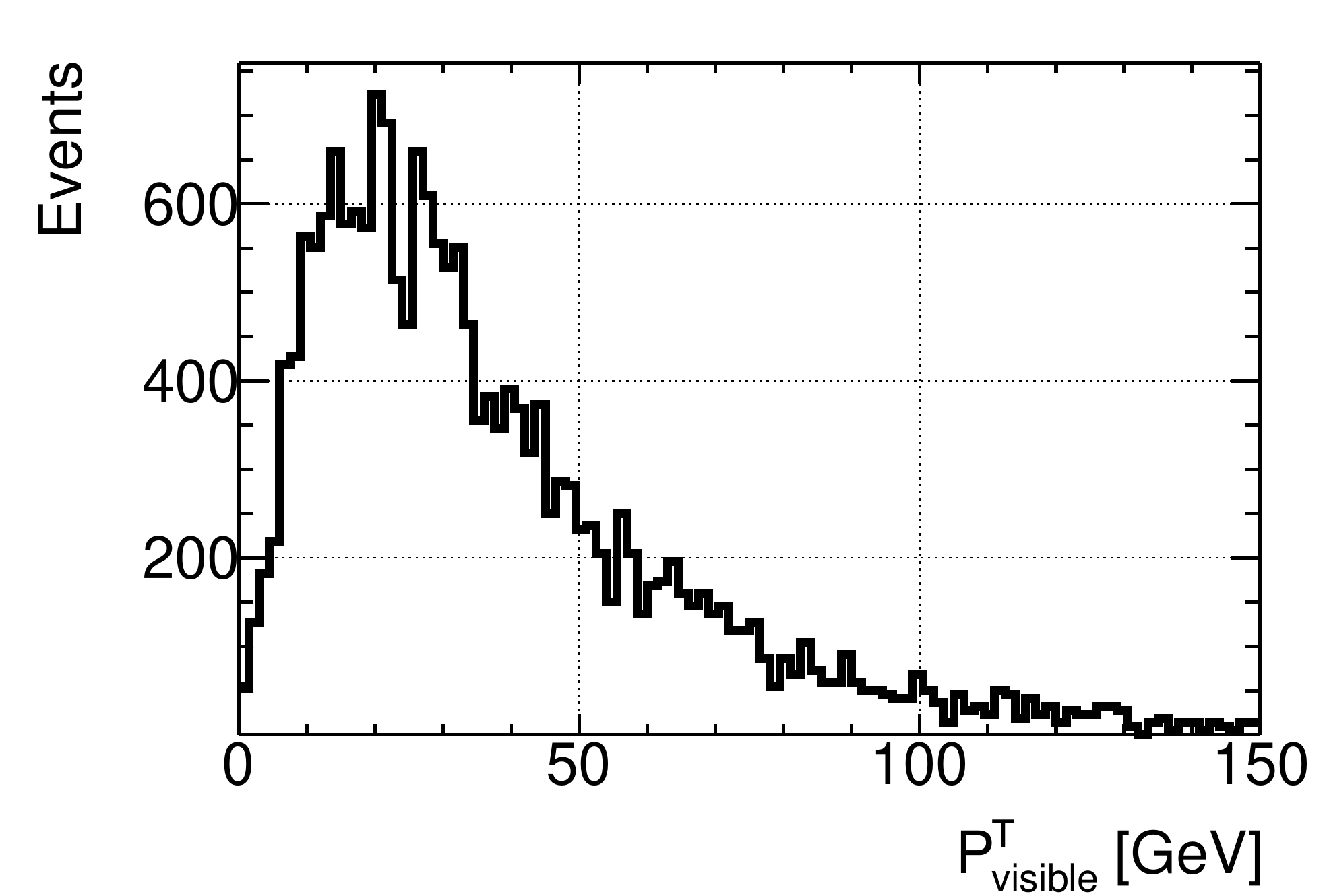} 
  \end{center}
  \caption{The MET (left) and the total transverse momentum of all visible particles (right) distributions for the BP4 signal.}\label{f_met_visible}
\end{figure} 

As demonstrated by  Figs.~\ref{f_mh1} and  \ref{f_mch}, even though there is some amount of combinatorics, the salient kinematic features of the signal stem clearly in a variety of mass  observables reconstructed from the $\ell\nu_\ell+4\gamma $ final state, which can then help to distinguish between signal and background events or else characterise the former, depending on the size of the latter.
For completeness, 
in Fig.~\ref{f_met_visible}, we show the MET distribution and total transverse momentum of all visible particles. The peaks of these two plots are at about 20 GeV, which means that one  could add a further MET cut to suppress possible backgrounds from the $B$ hadron decays, though  this should  not be
necessary, as we shall demonstrate next.

\subsection{Significances}

\begin{figure}[htp]
\begin{minipage}{0.47\textwidth}
    \begin{center}    
 \includegraphics[height=6cm]{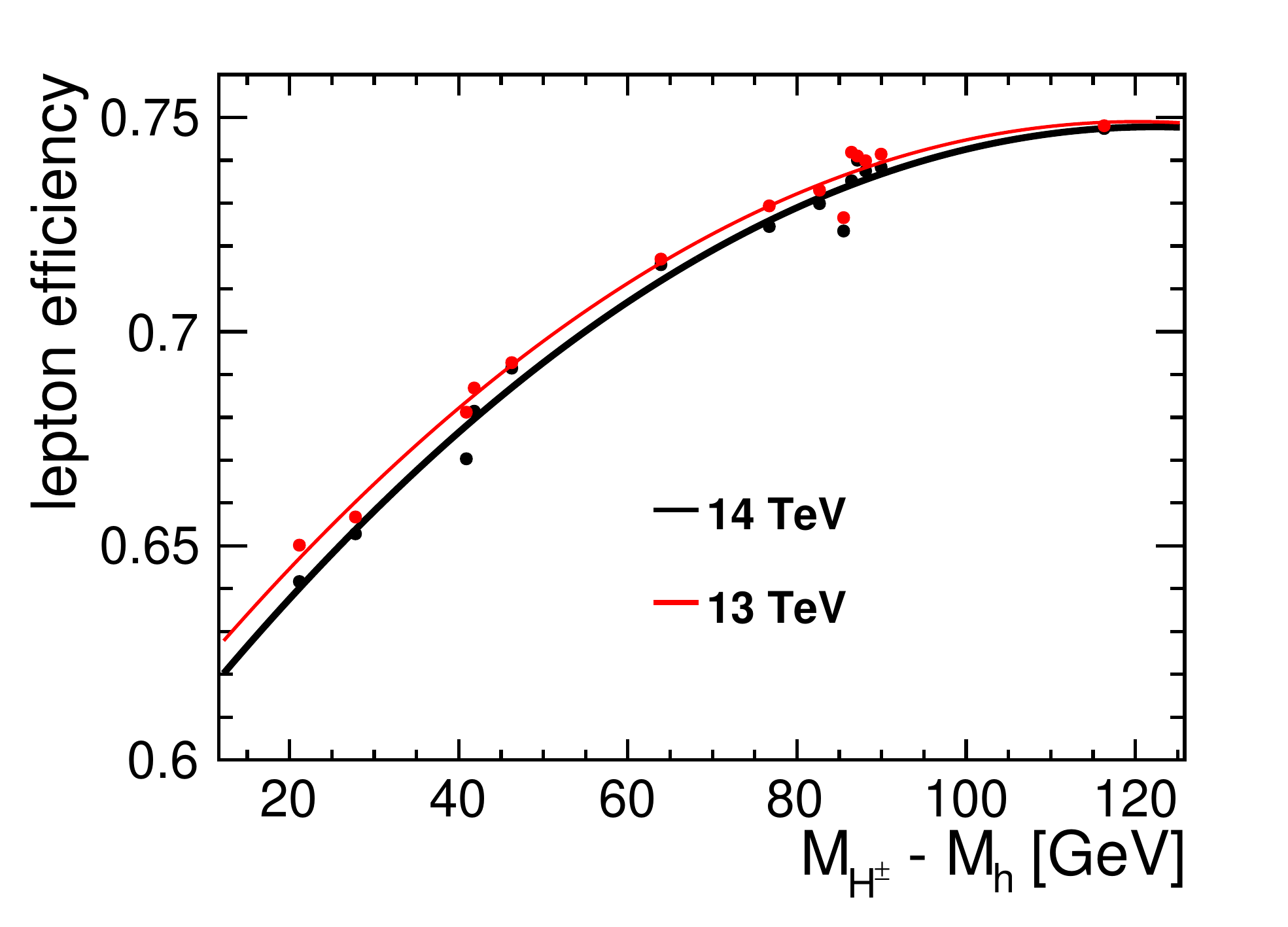} 
  \end{center}
 \end{minipage}
 \begin{minipage}{0.47\textwidth}
    \begin{center}    
 \includegraphics[height=6cm]{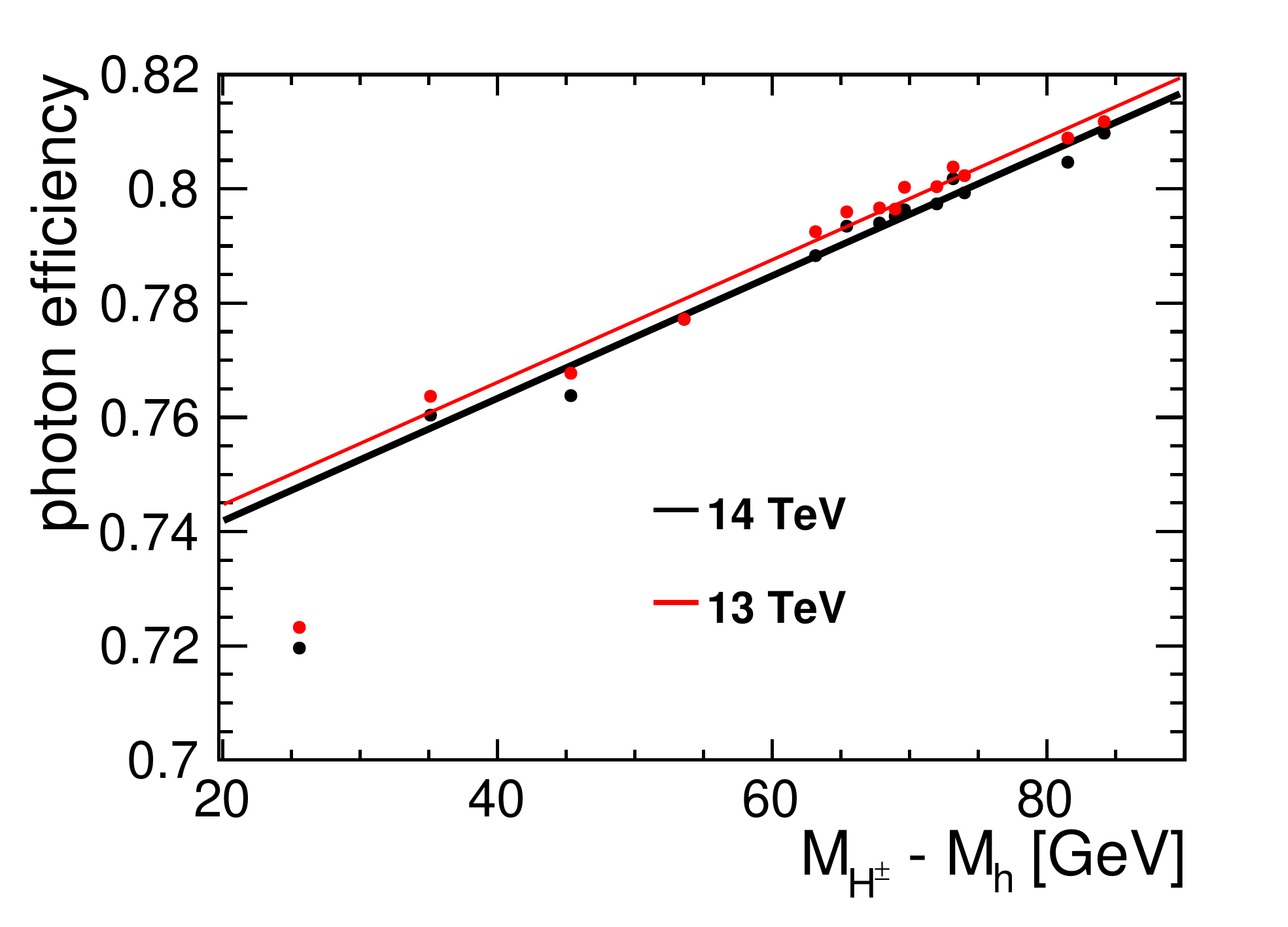} 
  \end{center}
 \end{minipage}
  \caption{The lepton tagging efficiency (left) and the photon tagging efficiency (right) are shown as function of the mass difference between the $H^\pm$ and $h$ states.
}\label{f_detector_effi}
\end{figure}

We now estimate the tagging efficiency for leptons and photons at detector level by using Delphes.  We generate 10k events for each BP and count the percentage of events where a lepton in the final state can be successfully reconstructed and recognised,  while for the photon case we count four photons total  efficiencies, then convert to single photon efficiency. The efficiencies for our BPs are s shown in Fig~\ref{f_detector_effi}.
By using some simple functions to fit the curves in Fig~\ref{f_detector_effi}, we find that the efficiencies for leptons and photons can be expressed by the following two relations, respectively \footnote{We should mention here that, to parameterise the typical photon efficiency, we used all points except BP1, which  has evidently a smaller efficiency that the other BPs, because $m_{h}$ is about 25 GeV, which in turn means that the rate for BP1  will be somewhat overestimated.}:
    \begin{align*}
\begin{split}
  \epsilon_{\ell}= \left \{
 \begin{array}{ll}
     0.59035+0.002574 x-1.0523\times 10^{-5} x^2, \text{ where } x=(M_{H^{\pm}}-M_{h})      & (14 ~\text{TeV}),\\
      0.59892+0.002500 x-1.0373\times 10^{-5} x^2, \text{ where } x=(M_{H^{\pm}}-M_{h})     & (13 ~\text{TeV}),\\
 \end{array}
 \right.
 \end{split}
 \end{align*}
     \begin{align*}
\begin{split}
  \epsilon_{\gamma}= \left \{
 \begin{array}{ll}
      0.001073\times M_{h}+0.72040,      & (14~ \text{TeV}),\\
       0.001072\times M_{h}+0.72327,     & (13~ \text{TeV}).\\
 \end{array}
 \right.
 \end{split}
 \end{align*}
Their shapes are captured in Fig.~\ref{f_detector_effi}.

Thus, we can derive the acceptance efficiency at detector level, $\epsilon_{\rm det}$, which can be expressed as a function of the mistagging rate of jets, lepton reconstruction and photon detection efficiencies as 
\begin{equation}
    \epsilon_{\rm det}=10^{-3\;n_{j}}\times \epsilon_{\ell}\times \epsilon_{\gamma}^{4-n_{j}},
\end{equation}
where $n_j$ denotes the jet number and $4 - n_{j}$ denotes the photon number. By using this acceptance efficiency at detector level, we can estimate the detection efficiency for the whole parameter space of our 	2HDM Type-I scenario.

Since the $W^\pm$ bosons in the final state could be either on-shell or off-shell, depending on the BP, in this work, we adopt two sets of cuts (see Ref.~\cite{Arhrib:2017wmo}) to examine how the efficiencies can change. The first set of cuts is
\bea
p_{T}^{\gamma} > 10 ~\textrm{GeV},~~~~ p_{T}^{\ell}>20 ~\textrm{GeV} \,, \label{set1}
\eea
and the second set of cuts is 
\bea
p_{T}^{\gamma} > 20 ~\textrm{GeV}, ~~~~p_{T}^{\ell}>10 ~\textrm{GeV}. \label{set2}
\eea
\begin{figure}[!t]
\begin{center}
    13 TeV \\
\end{center}
\begin{minipage}{0.47\textwidth}
    \begin{center}    
 \includegraphics[height=4.7cm]{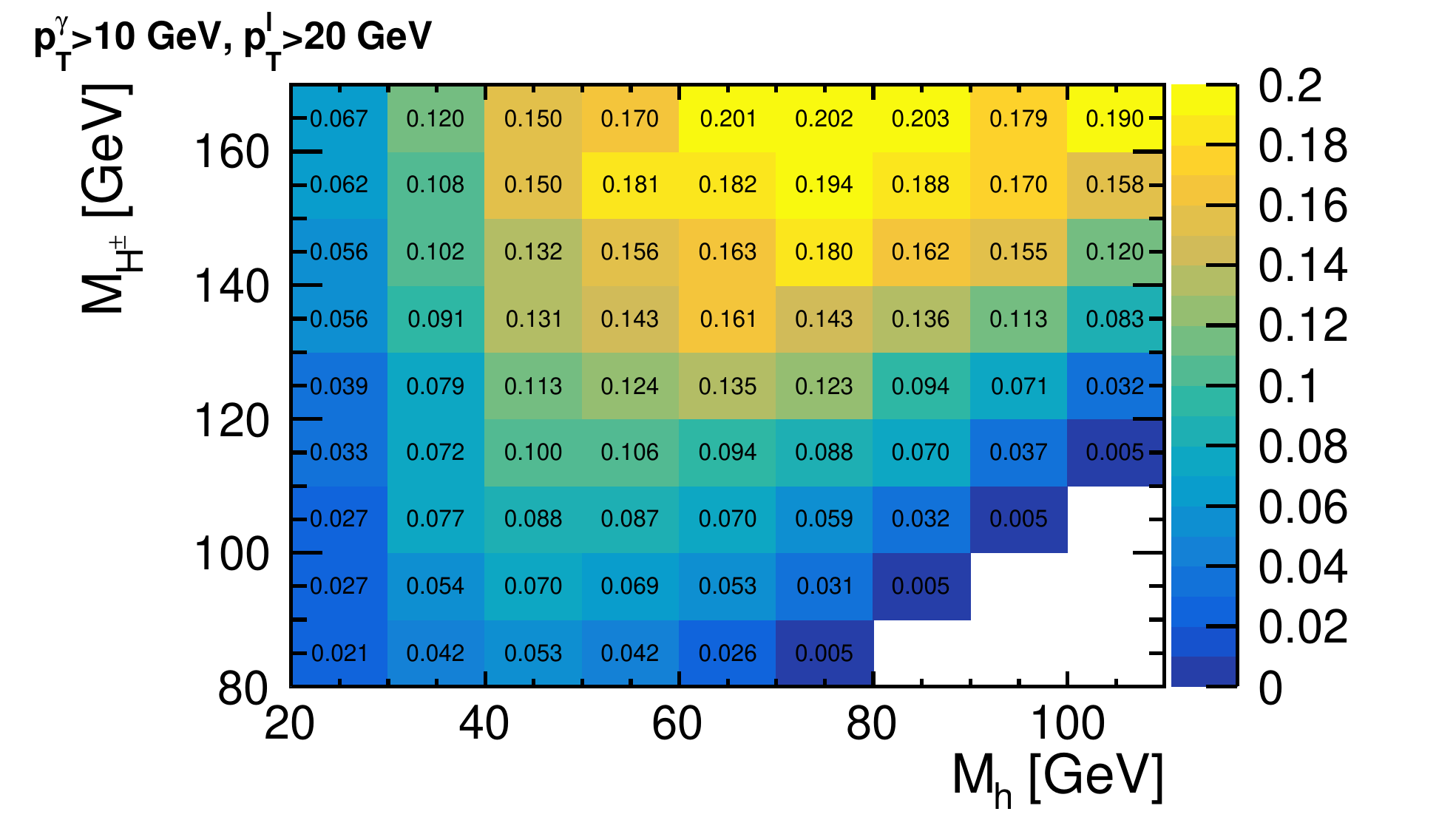} 
  \end{center}
 \end{minipage}
 \begin{minipage}{0.47\textwidth}
    \begin{center}    
 \includegraphics[height=4.7cm]{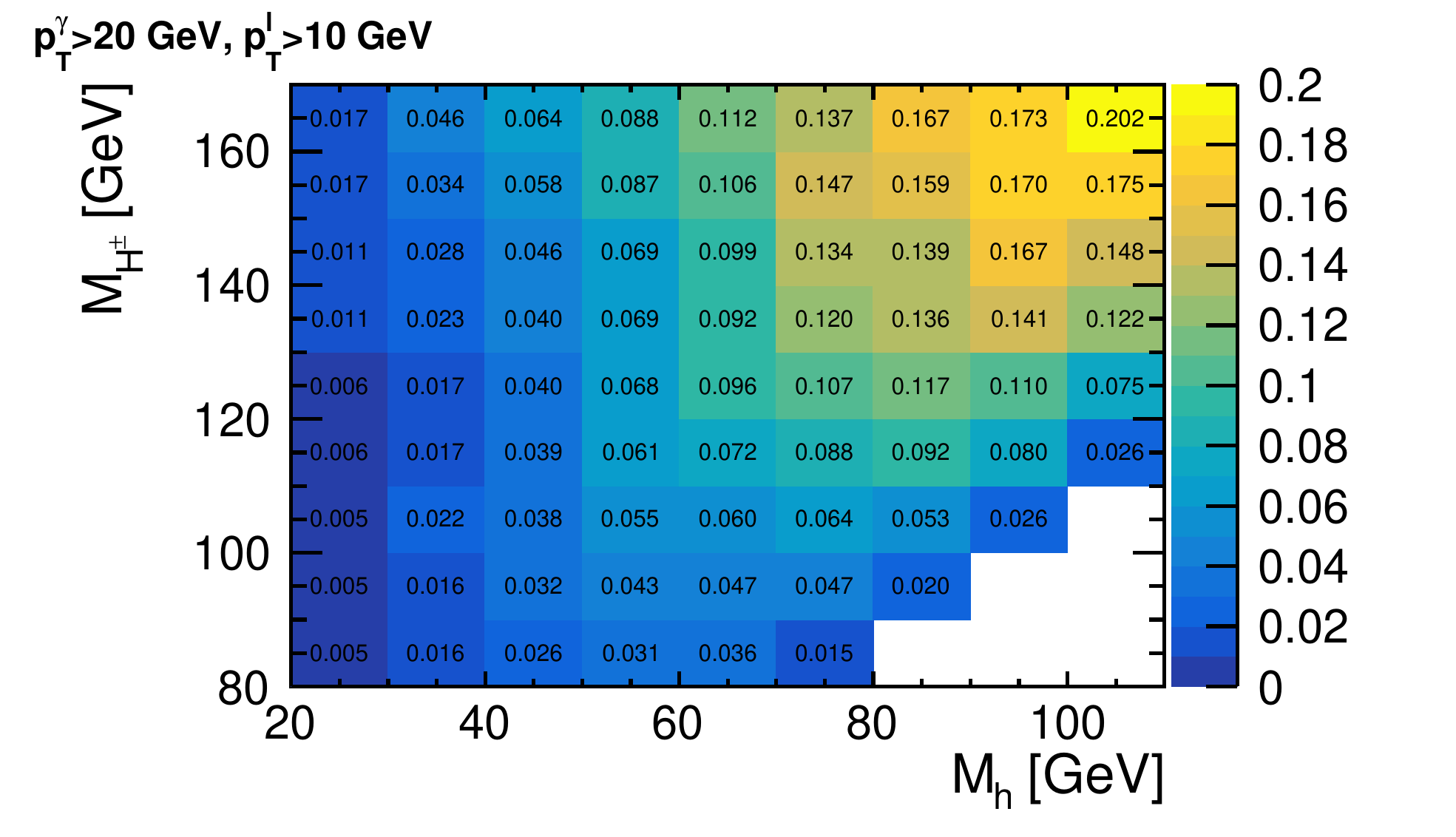} 
  \end{center}
 \end{minipage}
\begin{center}
    14 TeV \\
\end{center}
 \begin{minipage}{0.47\textwidth}
    \begin{center}    
 \includegraphics[height=4.7cm]{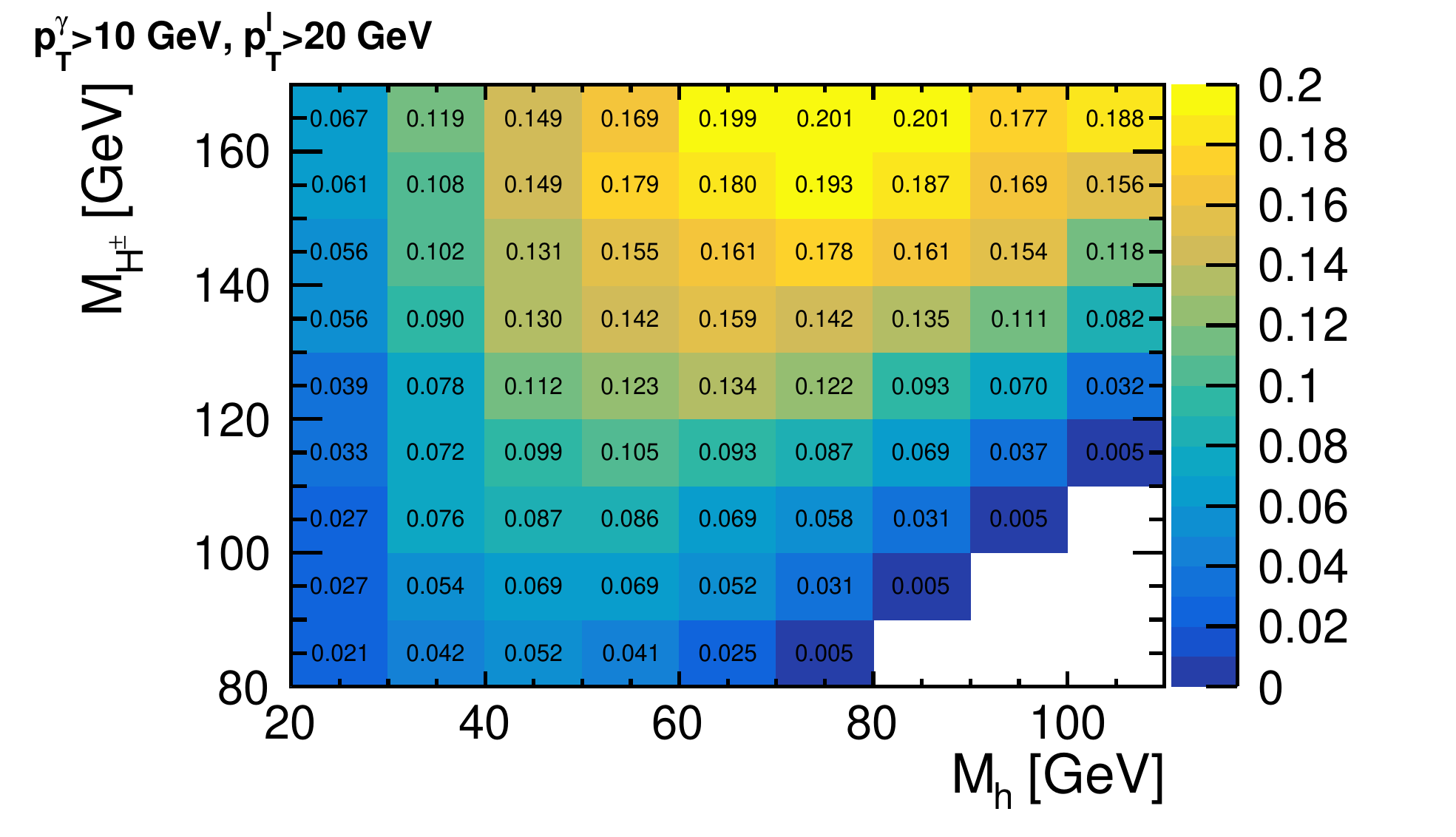} 
  \end{center}
 \end{minipage}
 \begin{minipage}{0.47\textwidth}
    \begin{center}    
 \includegraphics[height=4.7cm]{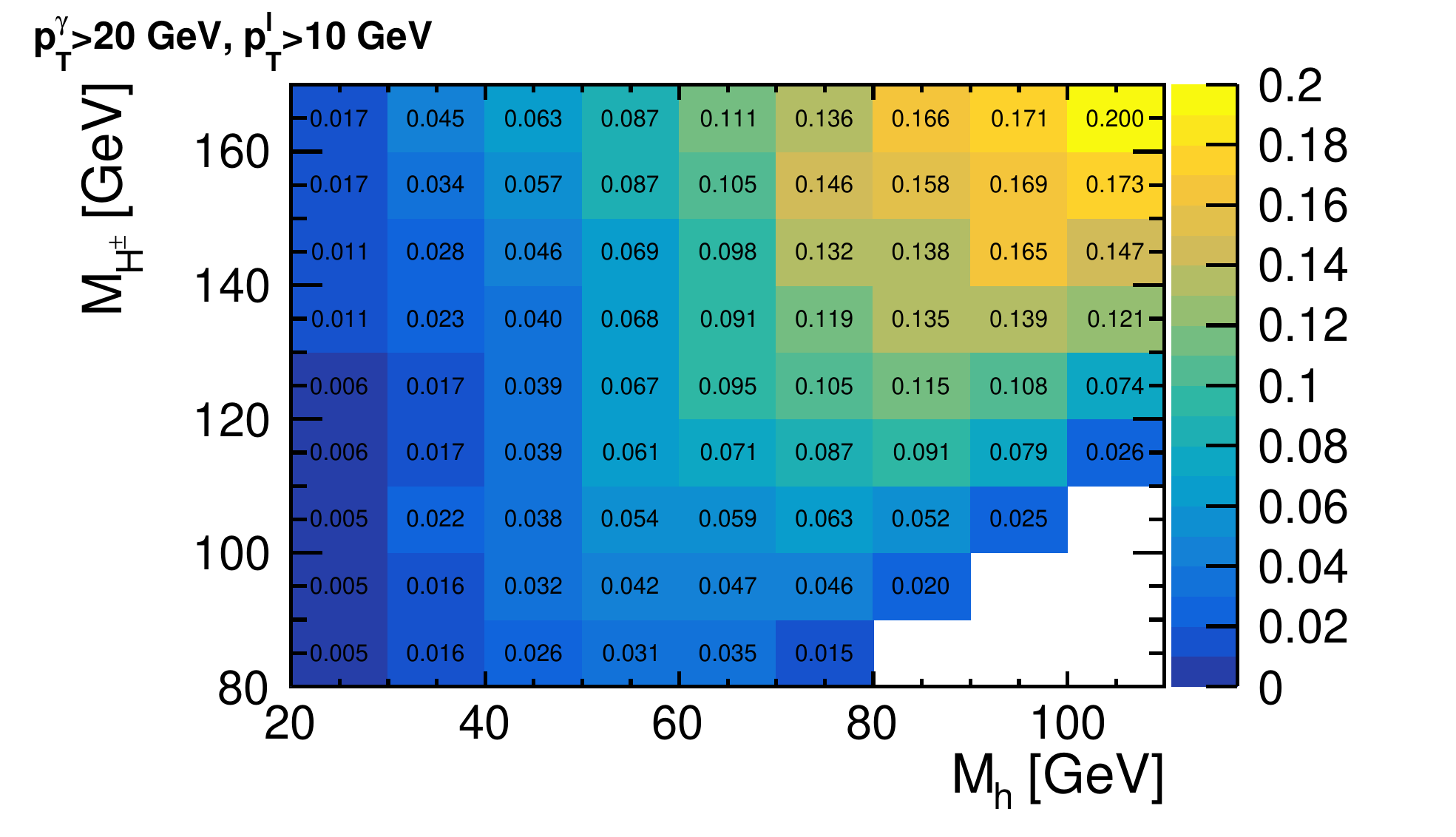} 
  \end{center}
 \end{minipage}
  \caption{Fiducial efficiency $\epsilon$ for detecting the signal via the  $\ell\nu_\ell+4\gamma$ signature at  detector level for the two sets of cuts given in Eqs.~(\ref{set1}) (left)  and (\ref{set2}) (right).}\label{f_effi}
\end{figure} 

To determine the fiducial efficiencies of each point in the parameter space, we use the relation \be
\epsilon= \sigma(\textrm{cuts}) \times \epsilon_{\rm det} / \sigma(\textrm{no cuts}).
\ee 
The results for the fiducial efficiencies are shown in Fig.~\ref{f_effi} for the two sets of cuts, which show a strong dependence on the masses of both charged and  neutral Higgs boson.  
Obviously, in the parameter region where the $W^\pm$ is on-shell and the $h$ not too light, a larger efficiency can be obtained. In contrast,  in the parameter space  region with a very light  $h$ (say $M_{h}$  around 25 GeV), the signal loss is caused by soft photons while,   in the parameter space  region with $M_{H^\pm} - M_{h}$  small, the signal loss is caused by soft leptons. When comparing our two sets of cuts, we can see that the first one has a better acceptance efficiency than the second one in covering a wider region of parameter space.

Before moving on to compute the significances of our signal for the BPs introduced, we present Tab.~\ref{t_crosssection_preselection_bk} for
the purpose of confirming the statement made in Ref.~\cite{Arhrib:2017wmo}, that none of the backgrounds is really observable for  any realistic LHC and HL-LHC luminosity.
(Results are shown here for 13 TeV, but the conclusion is the same for 14 TeV.)  We also present 
the predicted cross sections for the signals emerging from the BPs  after taking into account the cuts and the detector acceptance efficiency in Tabs.~\ref{t_crosssection_preselection_13} and ~\ref{t_crosssection_preselection_14}, where we have considered $\sqrt{s}=13$ and $14$ TeV, respectively. 
Due to the fact that the quark fluxes cannot be greatly enhanced when the collision energy increases from $13$ to $14$ TeV, we notice that the cross sections of the signal processes can only increase by $5\%$ to $10\%$ between the lower and higher center-of-mass energies.

\begin{table}
 \begin{center}
 \begin{tabular}{|c| c| c| c|}
 \hline
  Process & Cross section (fb)   &  After selection \\
  \hline
$W^\pm+4j0\gamma$ & 145890 & 0 \\
  \hline
$W^\pm+3j1\gamma$ & 1730 & 0 \\ 
  \hline
$W^\pm+2j2\gamma$ & 10.2 & $2.55\times 10^{-4}$\\ 
  \hline
$W^\pm+1j3\gamma$ & 0.0282 & $1.52\times 10^{-4}$\\ 
  \hline
$W^\pm+0j4\gamma$ & $1.69\times 10^{-5}$ & $5.71 \times 10^{-6}$\\
\hline
 \end{tabular}
 \end{center}
      \caption{The cross sections of background processes with $\sqrt{s}=13$ TeV are given, after taking into account  cuts and detector effects.}  \label{t_crosssection_preselection_bk}
\end{table}

\begin{table}
 \begin{center}
 \begin{tabular}{|c| c| c| c|}
 \hline
  Cross section (fb) &    MG  &  After selection&  Estimate    \\
  \hline
BP1&	2.09& 	0.42& 	0.49 \\  
  \hline
BP2&	7.43& 	1.89& 	1.87 \\
  \hline
BP3&	0.57& 	0.15& 	0.15 \\
  \hline
BP4&	1.17 &	0.29 &	0.30 \\
  \hline
BP5&	9.49 &	2.57 &	2.55 \\
  \hline
BP6&	12.48& 	3.65 &	3.60 \\
  \hline
BP7&	10.42& 	2.90 &	2.90 \\
  \hline
BP8&	8.09 &	2.18 &	2.16 \\
  \hline
BP9&	7.31 &	1.98& 	1.95 \\
  \hline
BP10&	4.72 &	1.47 &	1.47 \\
  \hline
BP11&	4.65 &	1.39 &	1.38 \\
  \hline
BP12&	4.59 &	1.39& 	1.39 \\
  \hline
BP13&	4.36 &	1.34& 	1.33 \\
  \hline
BP14&	4.23 &	1.34 &	1.35 \\
\hline
 \end{tabular}
 \end{center}
      \caption{The cross sections of our signal processes with $\sqrt{s}=13$ TeV are given, after taking into account cuts and  detector effects (we also show the results from our analytical estimate). } \label{t_crosssection_preselection_13}
\end{table}

\begin{table}
 \begin{center}
 \begin{tabular}{|c| c| c| c|}
 \hline
  Cross section (fb) &    MG  &  After selection&  Estimate    \\
\hline
BP1&   2.27&	0.44&	0.52 \\
  \hline
BP2&   8.02&	2.00&	1.99 \\
  \hline
BP3&   0.62&	0.15&	0.16 \\
  \hline
BP4&	1.27& 	0.31& 	0.32 \\
  \hline
BP5&	10.19& 	2.68& 	2.68  \\
  \hline
BP6&	13.46& 	3.86 &	3.81  \\
  \hline
BP7&	11.26& 	3.10& 	3.07  \\
  \hline
BP8&	8.76 &	2.30 &	2.29  \\
  \hline
BP9&	7.87& 	2.09 &	2.05  \\
  \hline
BP10&	5.16 &	1.59 & 	1.58  \\
  \hline
BP11&	5.03 &	1.48 & 	1.47  \\
  \hline
BP12&	4.93 &	1.47 & 	1.46  \\
  \hline
BP13&	4.76 &	1.44 & 	1.43  \\
  \hline
BP14&	4.61 &	1.43 & 	1.44  \\
  \hline
 \end{tabular}
 \end{center}
      \caption{The cross sections of our signal processes with $\sqrt{s}=14$ TeV are given, after taking into account cuts and  detector effects (we also show the results from our analytical estimate). } \label{t_crosssection_preselection_14}
\end{table}

\begin{table}
 \begin{center}
 \begin{tabular}{|c | c| c| c| c| c| c| c| c| c| c| c| c| c| c|}
 \hline
 BPs & 1 &   2 & 3 &   4  &  5 & 6 & 7 & 8 & 9 & 10 & 11 & 12 & 13 & 14   \\
  \hline
 $\sigma_{13 \text{TeV}}$ &12.1& 23.7& 6.7 & 9.4 & 27.4 & 32.6 & 29.2 & 25.2 & 23.9 & 20.8 & 20.2 & 20.3 & 19.9 & 19.9 \\
   \hline
 $\sigma_{14 \text{TeV}}$& 12.5& 24.4 &7.0 & 9.8 & 28.4 & 33.9 & 30.3 & 26.2 & 24.8 & 21.8 & 21.1 & 21.0 & 20.8 & 20.8 \\
  \hline
 \end{tabular}
 \end{center}
      \caption{The significances for all 14 BPs at the LHC are tabulated, where the luminosity is assumed to be 300 fb$^{-1}$ at both $\sqrt{s} =13$  and $14$ TeV. } \label{t_significance}
\end{table}

\begin{figure}[ht!]
\begin{center}
    13 TeV \\
\end{center}
\begin{minipage}{0.47\textwidth}
    \begin{center}    
 \includegraphics[height=4.7cm]{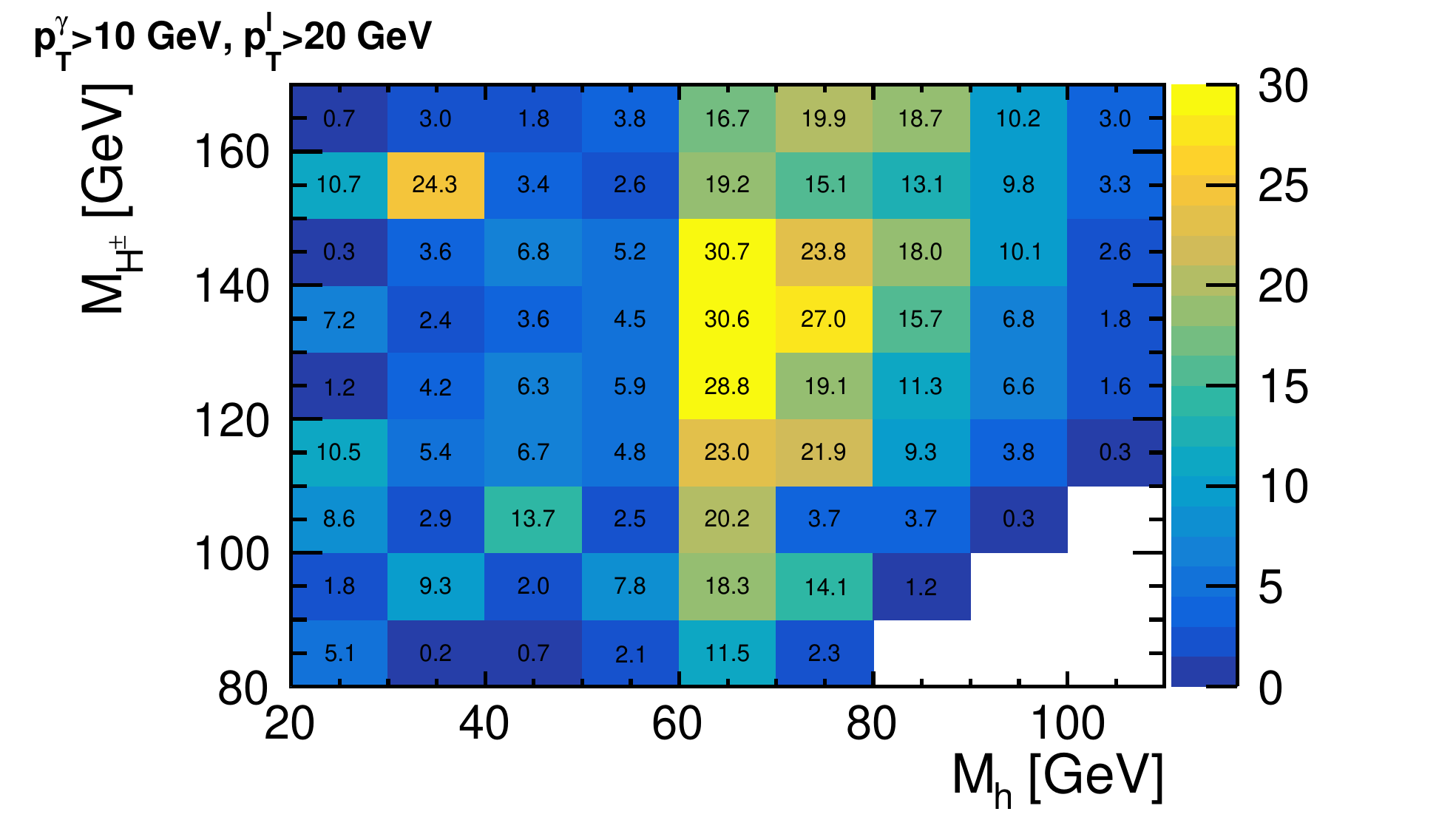} 
  \end{center}
 \end{minipage}
 \begin{minipage}{0.47\textwidth}
    \begin{center}    
 \includegraphics[height=4.7cm]{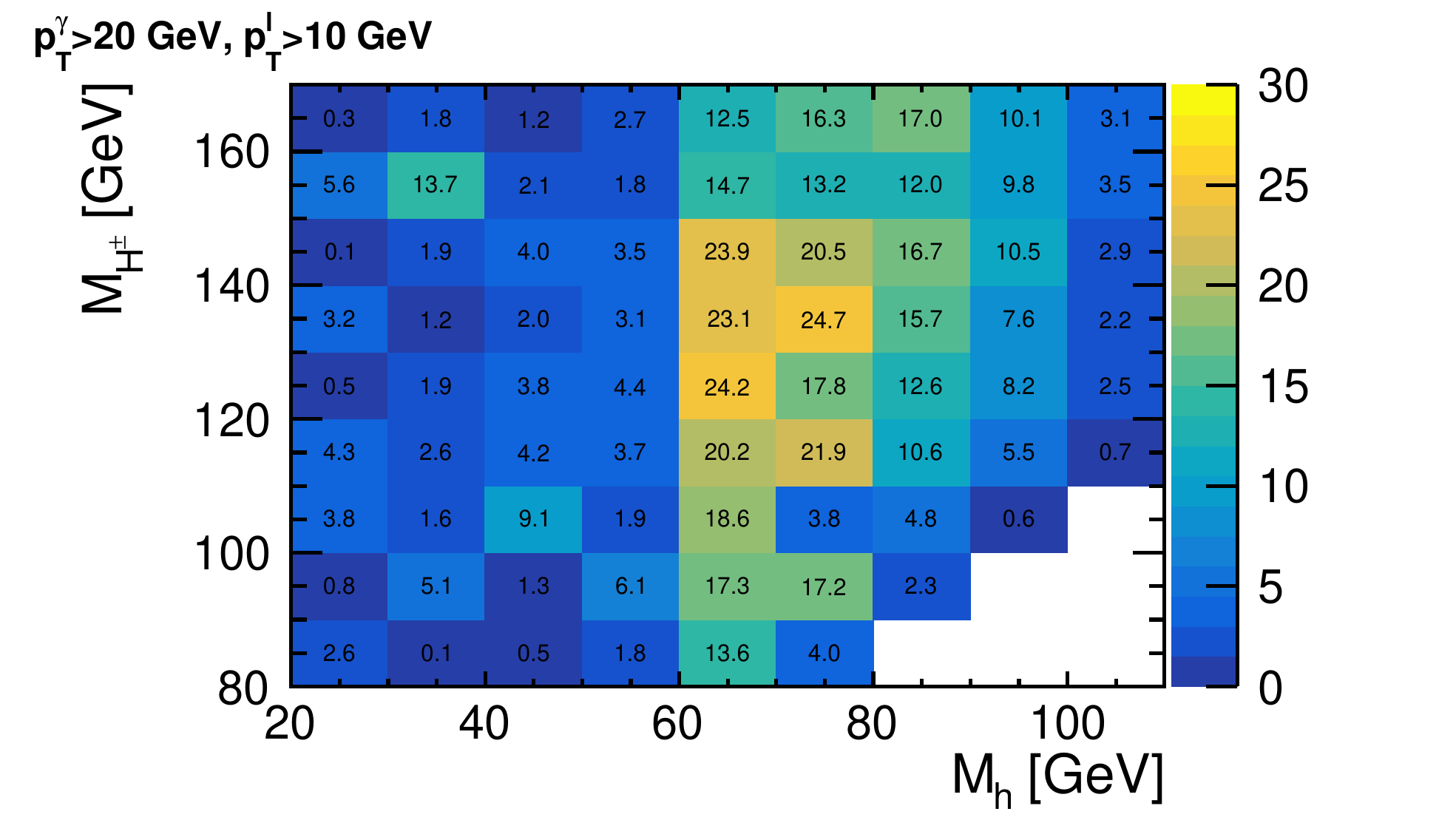} 
  \end{center}
 \end{minipage}
\begin{center}
    14 TeV \\
\end{center}
 \begin{minipage}{0.47\textwidth}
    \begin{center}    
 \includegraphics[height=4.7cm]{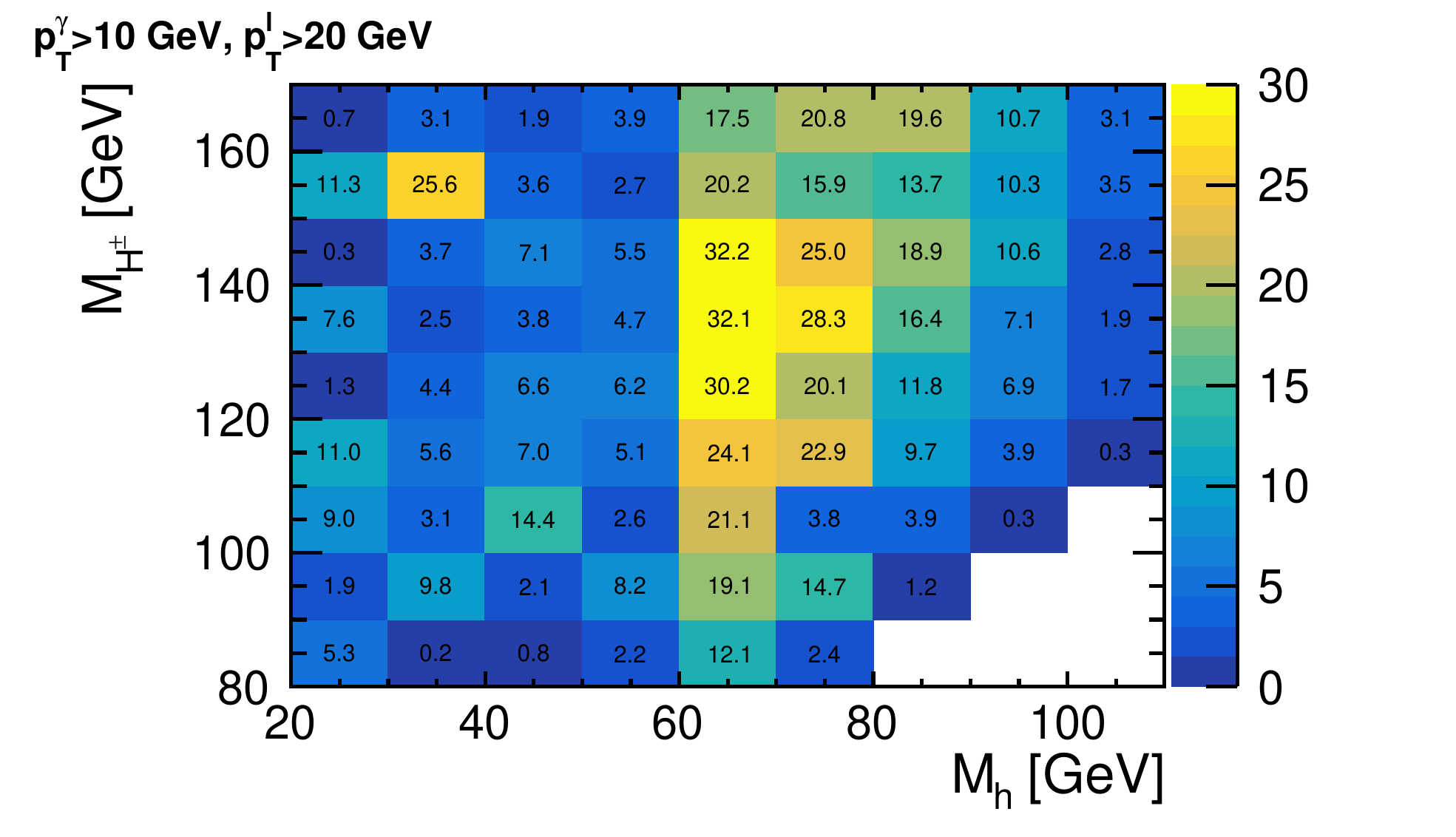} 
  \end{center}
 \end{minipage}
 \begin{minipage}{0.47\textwidth}
    \begin{center}    
 \includegraphics[height=4.7cm]{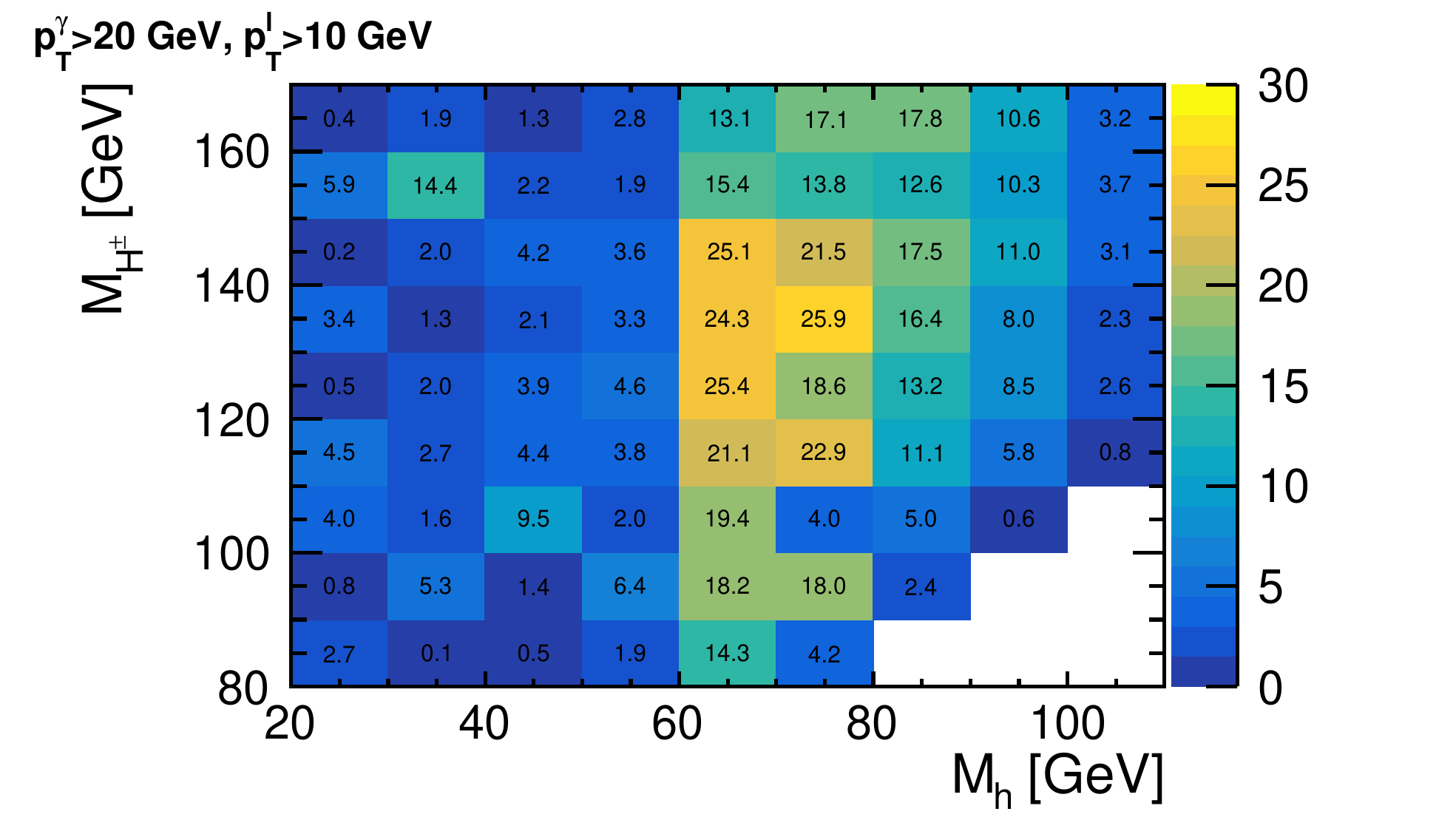} 
  \end{center}
 \end{minipage}
   \caption{The predicted significances over the  ($M_h, M_{H^\pm}$) plane for the two sets of cuts given in Eqs. (\ref{set1}) (left) and (\ref{set2}) (right),  when $\sqrt{s}=13$ TeV (top) as well as  $\sqrt{s}=14$ TeV (bottom) and $L=$ 300 fb$^{-1}$.}\label{f_sig_13}
\end{figure}

To compute the significances, due to the tiny number of background events, we can neglect the latter safely. Therefore, the predicted significances can be computed by  using the relation $\frac{N_{S}}{\sqrt{N_{S}+N_{B}}}\sim\sqrt{N_{S}}$. The corresponding results are shown in Tab. \ref{t_significance}. We find that the predicted significances are larger than 5 for all {14 BPs} in our 2HDM Type-I scenario when the luminosity is assumed to be 300 fb$^{-1}$, both  at $\sqrt{s}=13$ and $14$ TeV.  The predicted significances for both energies and the given luminosity over the $(M_h,M_{H^\pm})$ plane are shown in Fig. \ref{f_sig_13}, which are obtained from the described convolution of production cross sections with cut and acceptance efficiencies at detector level. To obtain the results given here, for each point  on the ($M_h,M_{H^\pm}$) plane, we allow $\tan\beta$ and $\sin(\beta-\alpha)$ to vary and take the maximal significance.  In such a figure, it should be pointed out that, when $m_{h} < 62.5$ GeV, the decay mode $H \to h h$ of the SM-like Higgs boson $H$ is open. Therefore, fewer  points in the parameter space are allowed by the Higgs boson data collected at the LHC, so that  the predicted significances drop when $M_{h}$ reaches the value of $62.5$ GeV. Another remarkable feature is that the predicted significances are larger than 5 for nearly all points on the ($M_h, M_{H^\pm}$) plane. In fact, only very few points on the grid have a significance below 2. Hence, in short, the whole interesting parameter space of our 2HDM Type-I  can be either ruled out or  discovered by the accumulated datasets of the entire Run 3 and/or only 1/10 of the HL-LHC one.

\begin{figure}[ht]
\begin{center}
    13 TeV \\
\end{center}
\begin{minipage}{0.47\textwidth}
    \begin{center}    
 \includegraphics[height=4.7cm]{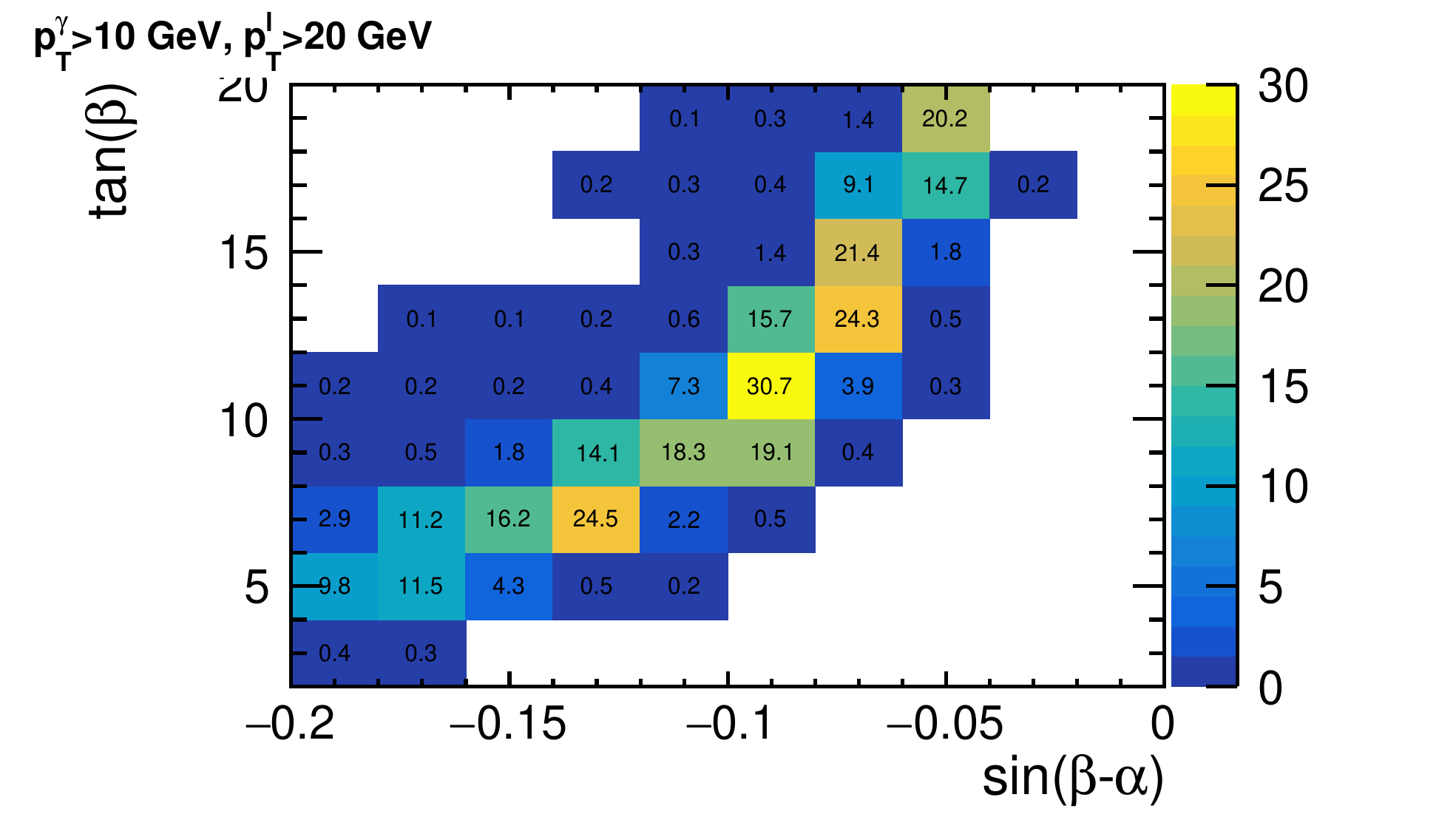} 
  \end{center}
 \end{minipage}
 \begin{minipage}{0.47\textwidth}
    \begin{center}    
 \includegraphics[height=4.7cm]{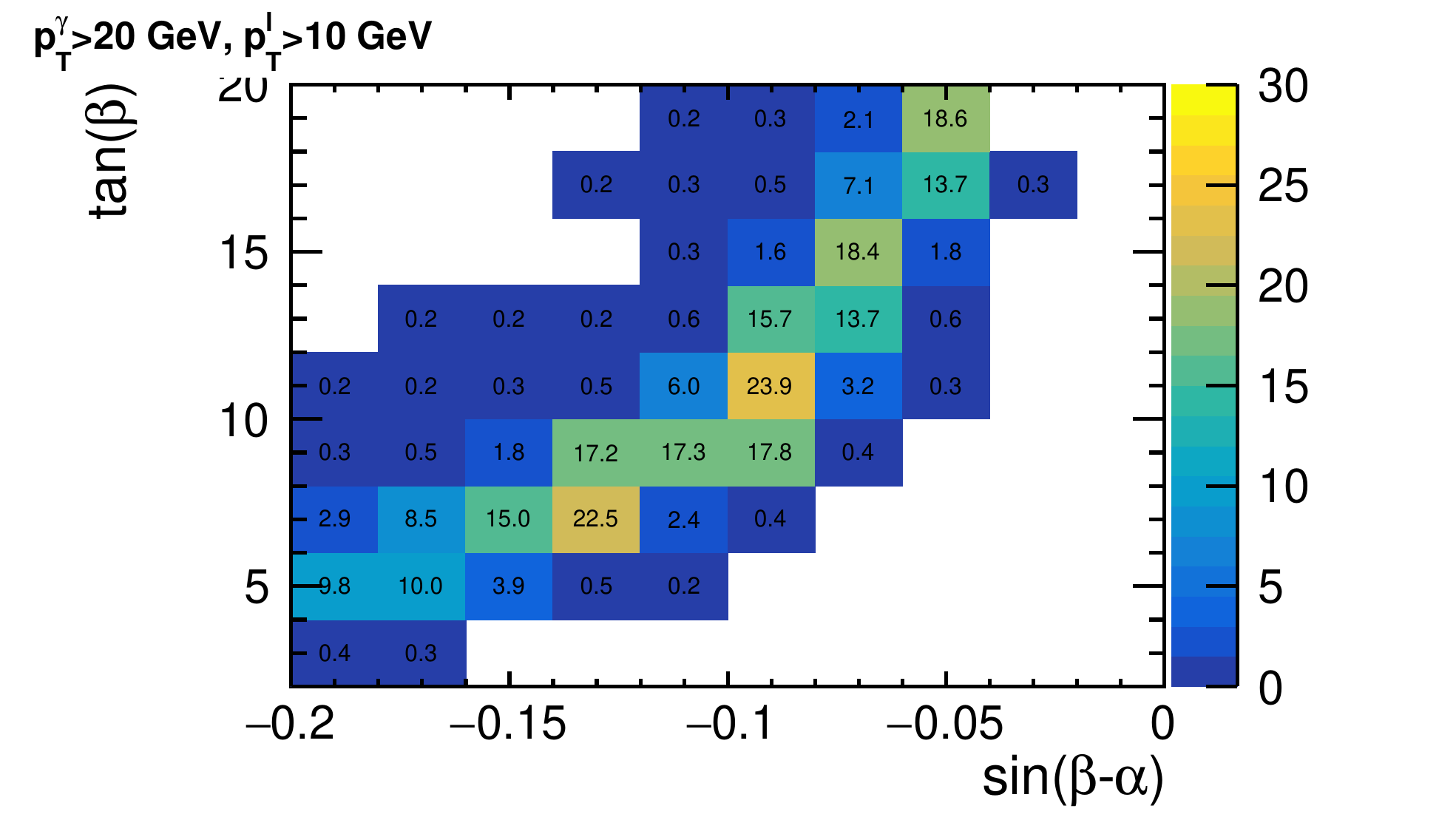} 
  \end{center}
 \end{minipage}
\begin{center}
    14 TeV \\
\end{center}
 \begin{minipage}{0.47\textwidth}
    \begin{center}    
 \includegraphics[height=4.7cm]{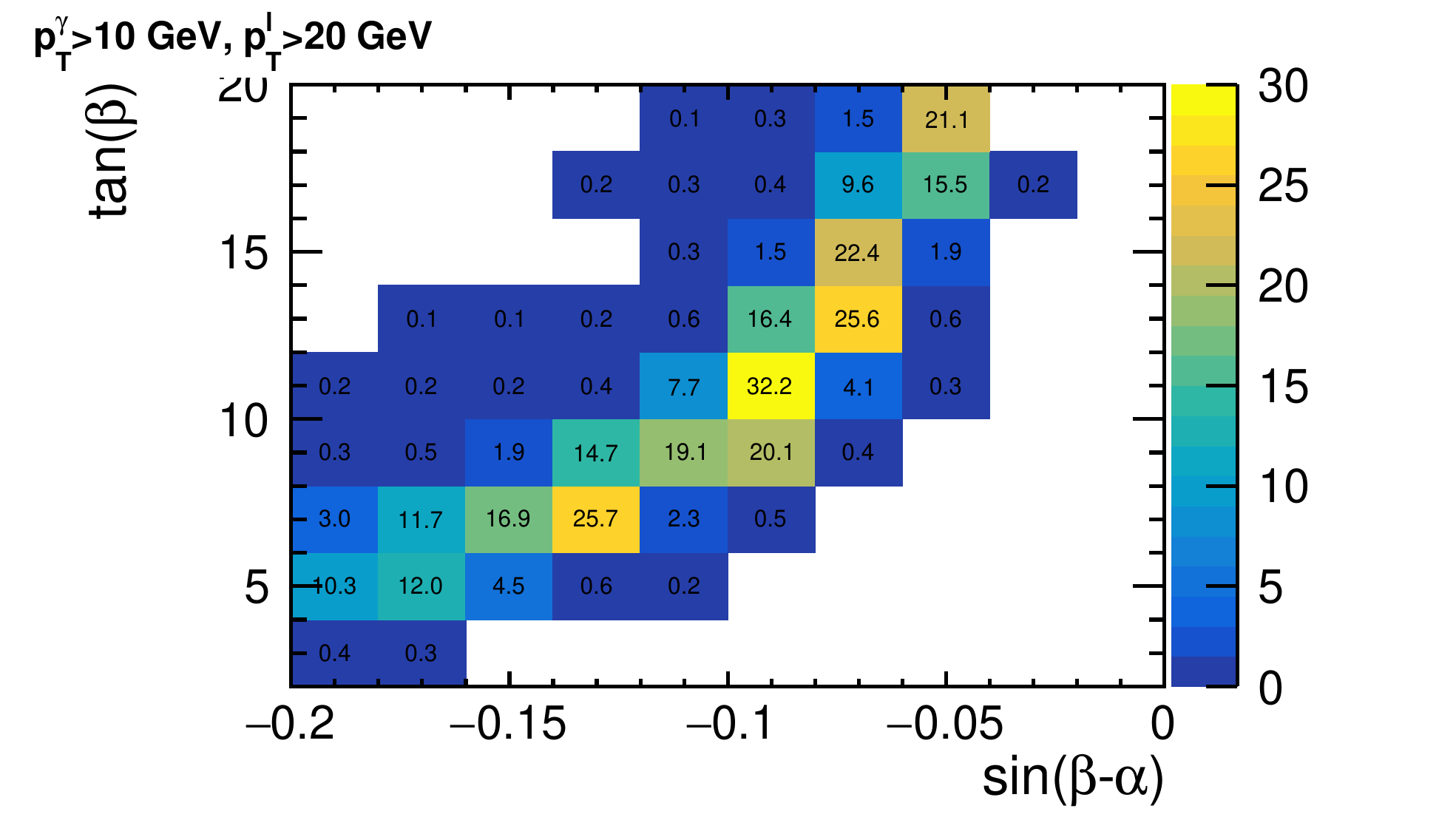} 
  \end{center}
 \end{minipage}
 \begin{minipage}{0.47\textwidth}
    \begin{center}    
 \includegraphics[height=4.7cm]{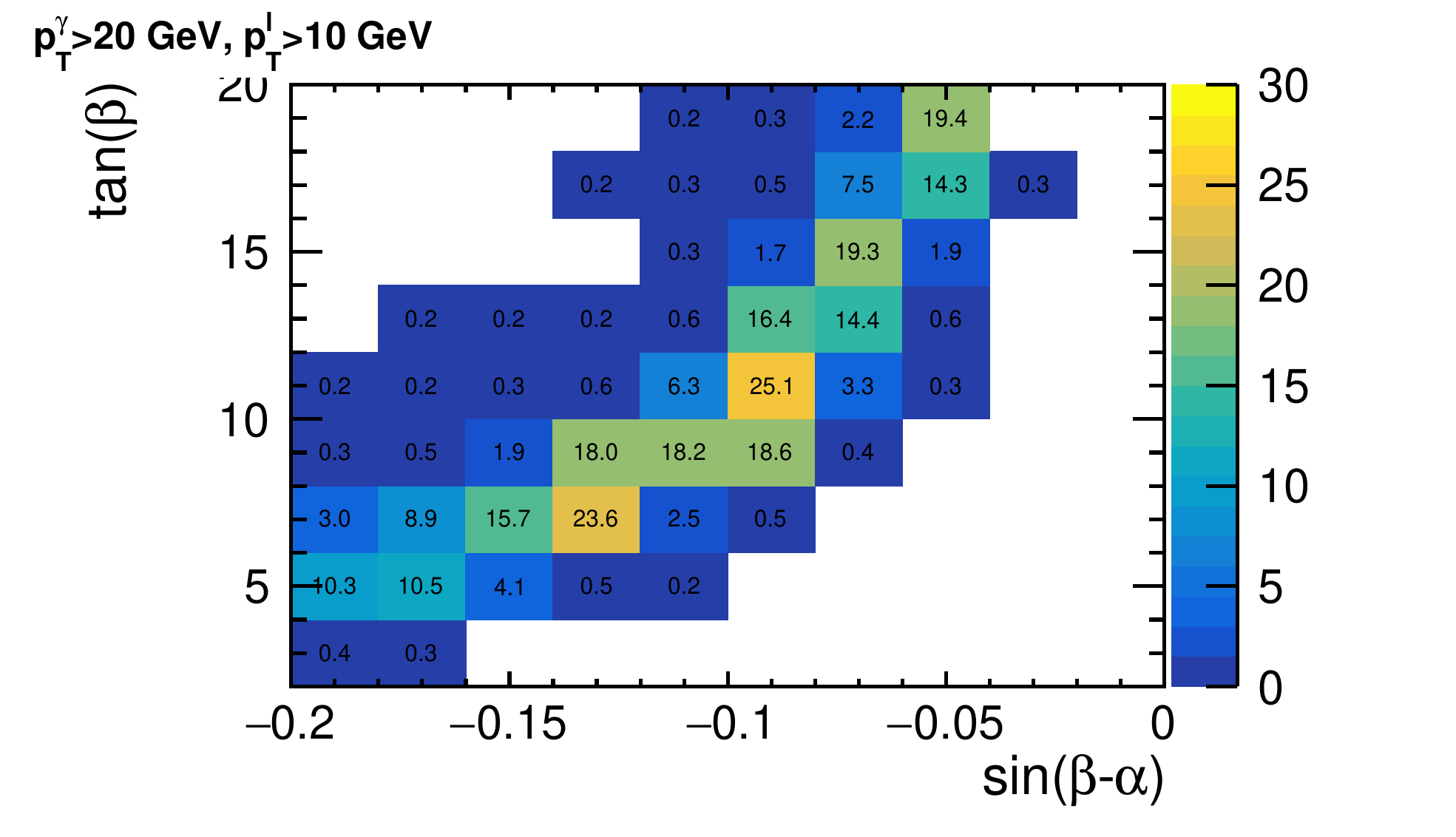} 
  \end{center}
 \end{minipage}
  \caption{The predicted significances over the ($\sin(\beta-\alpha), \tan\beta$)  plane for the two sets of cuts given in Eqs. (\ref{set1}) (left) and (\ref{set2}) (right), when $\sqrt{s}=13$ TeV (top) as well as  $\sqrt{s}=14$ TeV (bottom) and $L=$ 300 fb$^{-1}$.}\label{f_para_sig_13}
\end{figure}

The predicted significances in term of the 2HDM Type-I parameter space mapped over the ($\sin(\beta-\alpha), \tan\beta$)  plane are shown in Fig. \ref{f_para_sig_13} for both $\sqrt s=13$ and 14 TeV, respectively. To obtain the results given here, for each point on the ($\sin(\beta-\alpha), \tan\beta$)  plane, we allow $M_{H^\pm}$ and $M_h$ to vary and take again the maximal significance. This mapping makes it clear that some amount of fine-tuning in $\sin(\beta-\alpha)$ and/or  $\tan\beta$  is necessary to obtain large significances. However, for any $\tan\beta>5$, there is always a choice of $\sin(\beta-\alpha)$ that allows one to make a definite statement at both the LHC stages considered on the portion of parameter space of the 2HDM Type-I that we have sampled.

We conclude our numerical analysis by noting that, 
altogether, as  we can observe from Figs. \ref{f_sig_13}--\ref{f_para_sig_13}, it is the first set of cuts given in Eq. (\ref{set1}) that  yields  better significances than the one in  Eq. (\ref{set2}). 

\section{Conclusions}\label{sec_conclusion}
In this paper, we have examined the feasibility of the signature $\wboson+4\gamma$, where the $W^\pm$ decays leptonically in electrons and muons, from the associated production of the  charged Higgs boson and lightest neutral Higgs state of the 2HDM Type-I (i.e., via $pp \to H^\pm h\to W^{\pm(*)}hh\to \ell\nu_\ell + 4\gamma$) at the LHC with a collision energy of $\sqrt{s}=13$ and $14~\text{TeV}$ and an integrated luminosity of $L= 300$ $\fbinv$. Our analysis has been a detector level study exploiting full MC event generation including parton shower, hadronisation and heavy flavour  decays. By doing so, we have
confirmed a previous study done solely at the parton level, as we have proven that, even in presence of background generated by both real and fake photons (from jets), the  signal is essentially  background free, so that significances only depend upon the signal cross sections and the collider  integrated luminosities.  We have also provided some  reliable estimates for the detector efficiency and associated heat maps which can expedite an estimate of the signal significance over the relevant 2HDM Type-I parameter space, which we deem useful for current LHC working groups. Finally, for more thorough experimental analyses, we have also published  14 BPs, where the 
$W^\pm$ boson can be either on-shell or off-shell, depending on the mass difference $M_{H^\pm}-M_{h}$.    
\section*{Acknowledgements}
 The work of AA, RB, MK and BM is supported by the Moroccan Ministry of Higher Education and
Scientific Research MESRSFC and CNRST Project PPR/2015/6. The work of SM is supported in
part through the NExT Institute and the STFC Consolidated Grant No. ST/L000296/1. Y. W.
is supported by the ‘Scientific Research Funding Project for Introduced High-level Talents’ of the
Inner Mongolia Normal University Grant No. 2019YJRC001,  and the scientific research
funding for introduced high-level talents of Inner Mongolia of China. Q.-S. Yan’s work is supported by the
Natural Science Foundation of China Grant No. 11875260.

\bibliography{2hdmw4photon}
\end{document}